\documentclass[english,aps,prl,onecolumn, superscriptaddress,showpacs, amsmath]{revtex4-1}
\usepackage{graphicx}
\usepackage{float}
\usepackage[colorlinks=true,linkcolor=blue,citecolor=blue]{hyperref}
\begin{document}
\title{Localization physics in graphene Moire superlattices}	
\author{Chandan Kumar}
\author{Saurabh Kumar Srivastav}
\author{Priyo Adhikary}
\author{Sumilan Banerjee}
\author{Tanmoy Das}
\author{Anindya Das}
\email{anindya@iisc.ac.in}
\affiliation{Department of Physics, Indian Institute of Science, Bangalore 560 012, India}
\begin{abstract}
Non-trivial Berry phase of graphene leads to unusual quantum correction to the conductivity. Berry phase of $\pi$ in single layer graphene (SLG) and $2\pi$ in bi-layer graphene (BLG) is expected to reveal weak anti-localization (WAL) and weak localization (WL), respectively. However, experimentally both WAL and WL have been observed in graphene devices depending on the strength of different scattering mechanisms. Graphene superlattice having multiple Dirac cones is expected to exhibit $\pi$ to $2\pi$ Berry phase transition from primary Dirac cone (PDC) to cloned Dirac cone (CDC). However, its effect on localization physics has not been explored yet. In this letter we present the magneto-conductance study in a hexagonal Boron-nitride (hBN)-graphene moir$\acute{e}$ superlattice. Our results reveal a transition from WAL at PDC to WL at CDC in a single device by tuning the Fermi energy. The transition is supported by the quantum oscillation measurements showing a shift of $\pi$ phase from PDC to CDC and corresponding theoretical calculation capturing the Berry phase transition. Thus, our studies on localization physics in graphene superlattice pave the way to understand the carrier dynamics at multiple Dirac cones.
\end{abstract}

\pacs{}

\maketitle
The correlations between electron wave functions lead to quantum interference corrections to Drude-Boltzmann conductivity\cite{altshuler1980magnetoresistance,bergmann1984weak}. An electron traversing through a diffusive system gets scattered by impurities in all possible directions and electron trajectories form a close loop after multiple scattering. The difference of phase acquired by the electronic wave functions in such close loops, often described by time reversal paths, is zero leading to enhanced back-scattering and thereby decrease in conductance. This phenomena is referred as weak localization (WL)\cite{licini1985weakly}. However, it is realized that if the spin-orbit interaction (SOI) of the system is sufficiently large, the quantum interference results in increase of conductance, a phenomenon known as weak anti-localization (WAL)\cite{savchenko1983antilocalization,schierholz2002weak,miller2003gate,nihei2006gate}. Regimes of WL and WAL are sensitive to different type of symmetry breaking as well as scattering mechanisms in conventional two dimensional systems and have been studied extensively \cite{smorchkova1997spin,schierholz2002weak,miller2003gate,nihei2006gate,grbic2008strong}.

The physics of localization in SLG is much richer due to its relativistic nature. The low energy excitation in SLG is described by relativistic Dirac spinors with two-component pseudospin. This additional pseudospin quantum number gives rise to $\pi$ and $2\pi$ Berry phase in SLG and BLG, respectively\cite{ando1998berry}. The Berry phase add additional phase correction to the quantum interference and it has been predicted theoretically that the SLG and BLG would manifest WAL and WL, respectively\cite{ando1998berry,suzuura2002crossover,mccann2006weak,nomura2007quantum,morpurgo2006intervalley,kechedzhi2007influence,khveshchenko2006electron}. However, experimentally both WL\cite{morozov2006strong,heersche2006bipolar,berger2006electronic,tikhonenko2009transition,berezovsky2010imaging,chen2010magnetoresistance,
baker2012weak,mahmood2013epitaxial,iqbal2014enhanced,hilke2014weak,chandni2015transport,pal2012direct, ki2008inelastic,tikhonenko2008weak,oh2010electronic,moser2010magnetotransport,lundeberg2010rippled} and WAL\cite{wu2007weak,tikhonenko2009transition,liu2011enhanced,jouault2011interplay} have been reported in SLG. The interplay between WL and WAL in SLG rely on the relative strength of different symmetry breaking processess\cite{morpurgo2006intervalley,mccann2006weak,kechedzhi2007influence,morozov2006strong,tikhonenko2008weak,fal2007weak,lundeberg2010rippled,liu2011enhanced,chen2011negative,chandni2015transport}.
In a clean SLG device WAL will dominate at Dirac point whereas the presence of strong inter valley scattering will restore WL. Thus, the physics of localization in SLG is very intricate and depends on type of scatterers\cite{morozov2006strong,khveshchenko2006electron,mccann2006weak,aleiner2006effect,nomura2007quantum} which varies from device to device. 

In this context, the graphene superlattice (GSL) is an ideal platform to study the effect of Berry phase on localization physics for the following reasons. In a GSL extra set of Dirac cones known as cloned Dirac cones (CDC) appear symmetrically around the primary Dirac cone (PDC). Very recently, it has been shown experimentally\cite{wang2015topological} that the Berry phase changes from $\pi$ to $2\pi$ from PDC to CDC. Motivated by this Berry phase transition\cite{wang2015topological} we have carried out magneto-conductance (MC) studies on a GSL having multiple Dirac cones. The GSL is created using hetero-structure of hexagonal boron nitride (hBN) and SLG, where the small angle mismatch between the crystallographic planes of SLG and hBN generates weak periodic moir$\acute{e}$ potential. The MC studies at small magnetic fields at different carrier concentrations ($n$) and temperatures show a negative MC around the PDC whereas it is positive around the CDC. The experimental data with the theoretical fitting suggests that the WAL dominates around the PDC whereas WL dominates around the CDC. In order to investigate the source of transition from WAL to WL we further pursue quantum oscillations measurements as a function of magnetic field, which clearly shows the transition of Berry phase from $\pi$ at PDC to $2\pi$ at CDC. These experimental observations with our corresponding theoretical framework unambiguously reveal the effect of Berry phase on localization physics in GSL.

The GSL heterostructure is prepared by making stack of hBN/SLG/hBN using pick up technique\cite{zomer2011transfer}. With this technique the SLG remains pristine as it is not exposed to any environmental residue. The schematic of the device is shown in Fig. 1(a), where the conductance of the device is measured between the source and drain using standard lock-in technique and the global back gate ($V_{BG}$) is used to change $n$. Gate voltage response of the device is shown in Fig. 1(b) at 2 K. Along with the resistance peak at charge neutrality point (CNP) near $V_{BG} \sim 0$ V two more resistance peaks are observed placed symmetrically around CNP at $V_{BG}=\pm 42$V. The resistance peak in the hole side is much stronger compared to the electron side is consistent with the previous reports\cite{yankowitz2012emergence,hunt2013massive,dean2013hofstadter,ponomarenko2013cloning,yu2014hierarchy,woods2014commensurate,gorbachev2014detecting,wang2015evidence,kumar2016tunability}. The energy separation between the PDC and CDC  is $\sim 190$ meV which corresponds to moir$\acute{e}$ wavelength of $\sim$12 nm.
	\begin{figure}[ht!]
		\includegraphics[width=0.48\textwidth]{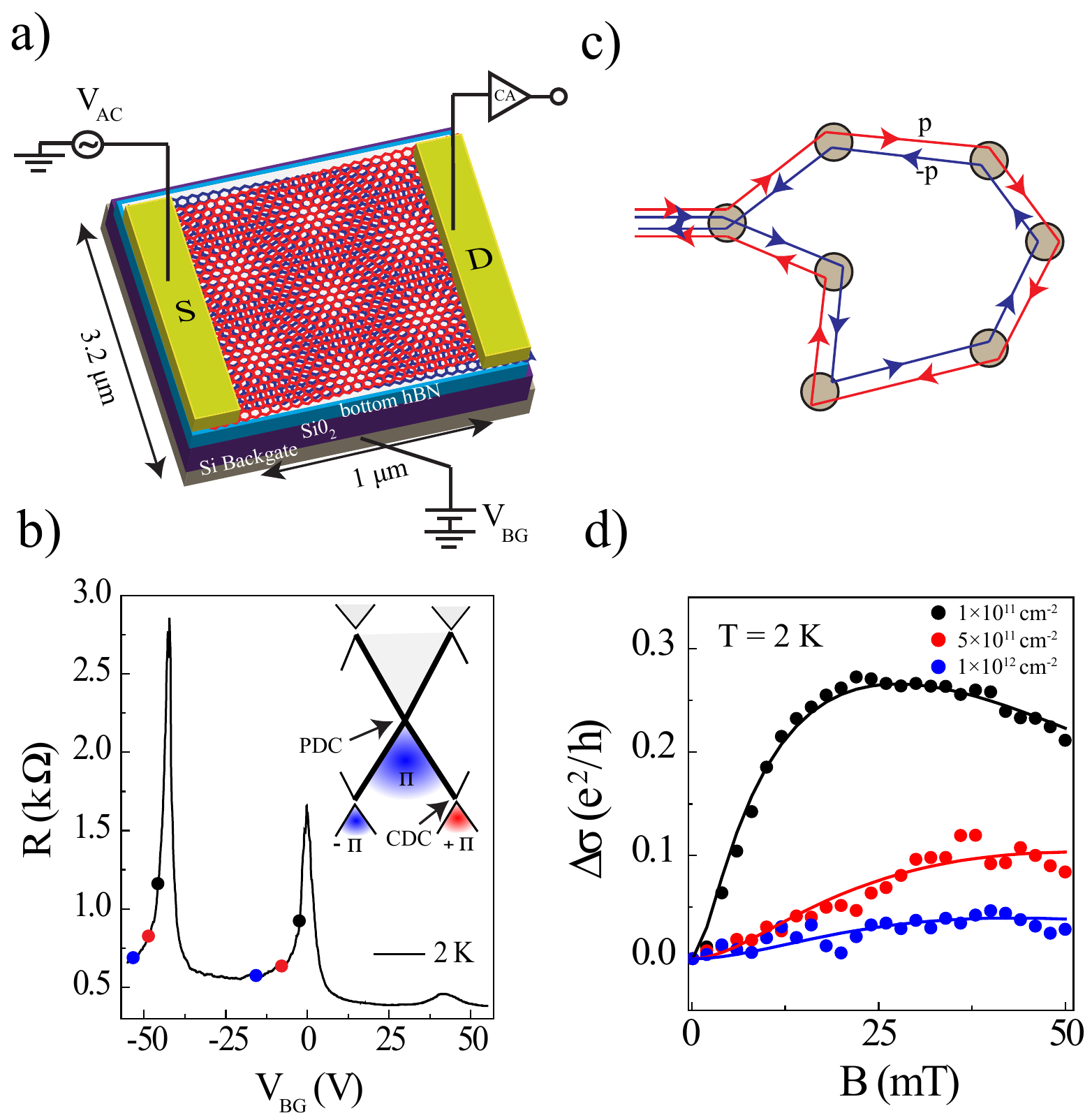}
		\caption{(a) The schematic of the device with the moir$\acute{e}$ superlattice potential created by hBN. The blue and the red hexagons represent the graphene and the hBN, respectively. (b) The resistance as a function of gate voltage at 2 K for zero magnetic field. Inset: Schematic diagram of band dispersion of graphene moir$\acute{e}$ superlattice. The Berry phase around a PDC is $\pi$. Although the Berry phase around the individual CDC points are $\pi$ with opposite sign but the total Berry phase for the electrons enclosing the entire superlattice Brillouin zone is zero ($2\pi$). (c) The schematic of the electron trajectories in presence of impurities that give rise to quantum correction to conductivity. (d) The magneto-conductance data for CDC at different carrier concentrations at 2 K. The solid curves are fit to Eq. (1).}
		\label{fig:image1}
	\end{figure}

In order to measure the effect of quantum interference on conductivity a small perpendicular magnetic field is applied which introduces an extra phase among the interfering electron wave functions as shown in Fig. 1 (c). As a result, the interference condition changes and conductance will increase (decrease) with magnetic field for WL (WAL). The MC in graphene is derived by MacCann \textit{et al.} \cite{mccann2006weak}

\begin{eqnarray}
\Delta\sigma(B)&=&\frac{e^{2}}{\pi h} \left[F\left(\frac{\tau_B^{-1}}
{\tau_{\phi}^{-1}}\right)-F\left(\frac{\tau_B^{-1}}{\tau_{\phi}^{-1}+
	2\tau_{i}^{-1}}\right)\right.\nonumber\\
&&-2F\left.\left(\frac{\tau_B^{-1}}{\tau_{\phi}^{-1}+
	\tau_{i}^{-1}+\tau_{*}^{-1}}\right)\right]\; .\label{eqn:one}
\end{eqnarray}
where $F (z) = ln(z)+\psi \left( \frac{1}{2} +\frac{1}{z}\right)$ , $\psi$ is the digamma function, $ \tau_{B}^{-1} = \frac{4eDB}{\hbar}$, $\tau_{\phi}^{-1}$ is the phase breaking rate, D is the diffusion coefficient, $\tau_{i}^{-1}$ is the intervalley scattering rate and $\tau_{*}^{-1}$ is the intravalley scattering rate. The main source of inter-valley scattering are the short range scatterers like the edge of the sample or sharp defects where as the dislocations, lattice defects and ripples are the sources of intra-valley scattering. In the case of clean sample $\tau_{i,*} \rightarrow \infty$, the first two terms in Eq. 1 cancel each other and the MC is governed by the third term, giving rise to WAL. In the opposite limit of strong inter-valley and intra-valley scattering (small $\tau_{i,*}$) the last two terms in Eq. (1) are suppressed and the first term gives WL.

Fig. 1 (d) shows the change of MC, $\Delta \sigma(B) = \sigma(B)-\sigma(B=0)>0$, with magnetic field at T = 2 K for different $n$ ($ 1\times10^{11} - 1\times10^{12}$ $cm^{-2}$) around the CDC as marked by the solid circles in Fig. 1 (b). The MC data around the CDC and PDC at different temperatures and $n$ (marked by the solid circles in Fig. 1b) are shown in Fig. 2. The solid curves in Fig. 2 are theoretical fitting curves with Eq. (1) to extract the different scattering rate (SI). With small magnetic field the increment of MC at CDC and decrement of MC at PDC suggest the signature of WL and WAL at CDC and PDC, respectively.

For small magnetic field the Eq. (1) can be reduced to\cite{tikhonenko2009transition}

\begin{eqnarray}
\Delta\sigma(B)&=&\frac{e^{2}}{24\pi h}\left(\frac{4eDB\tau_{\phi}}{\hbar}\right)^{2}\left[1- \frac{1}
{(1+2\tau_{\phi}/\tau_i)^2}\right.\nonumber\\
&&\left. - \frac{2}{(1+\tau_{\phi}/\tau_i+\tau_{\phi}/\tau_{*})^2}
\right]\; .\label{eqn:two}
\end{eqnarray}
and the sign of Eq. (2) determines WL or WAL. $\Delta \sigma(B)=0$ curve obtained from Eq. (2) is shown by a solid curve in Fig. 3 as a function of $\tau_{\phi}/\tau_{*}$ and $\tau_{\phi}/\tau_{i}$. This curve separates the region between WL and WAL. The fitted values of $\tau_{i}$, $\tau_{*}$ and $\tau_{\phi}$ from Fig. 1 (d) and Fig.2 (SI) are used to generate the data points in Fig. 3 (a) and Fig. 3 (b) for CDC and PDC, respectively. It can be clearly seen from Fig. 3 that most of the data points lie in WL and WAL region for CDC and PDC, respectively even for one order variation of temperature and density.

	\begin{figure}[ht!]
		\includegraphics[width=.48\textwidth]{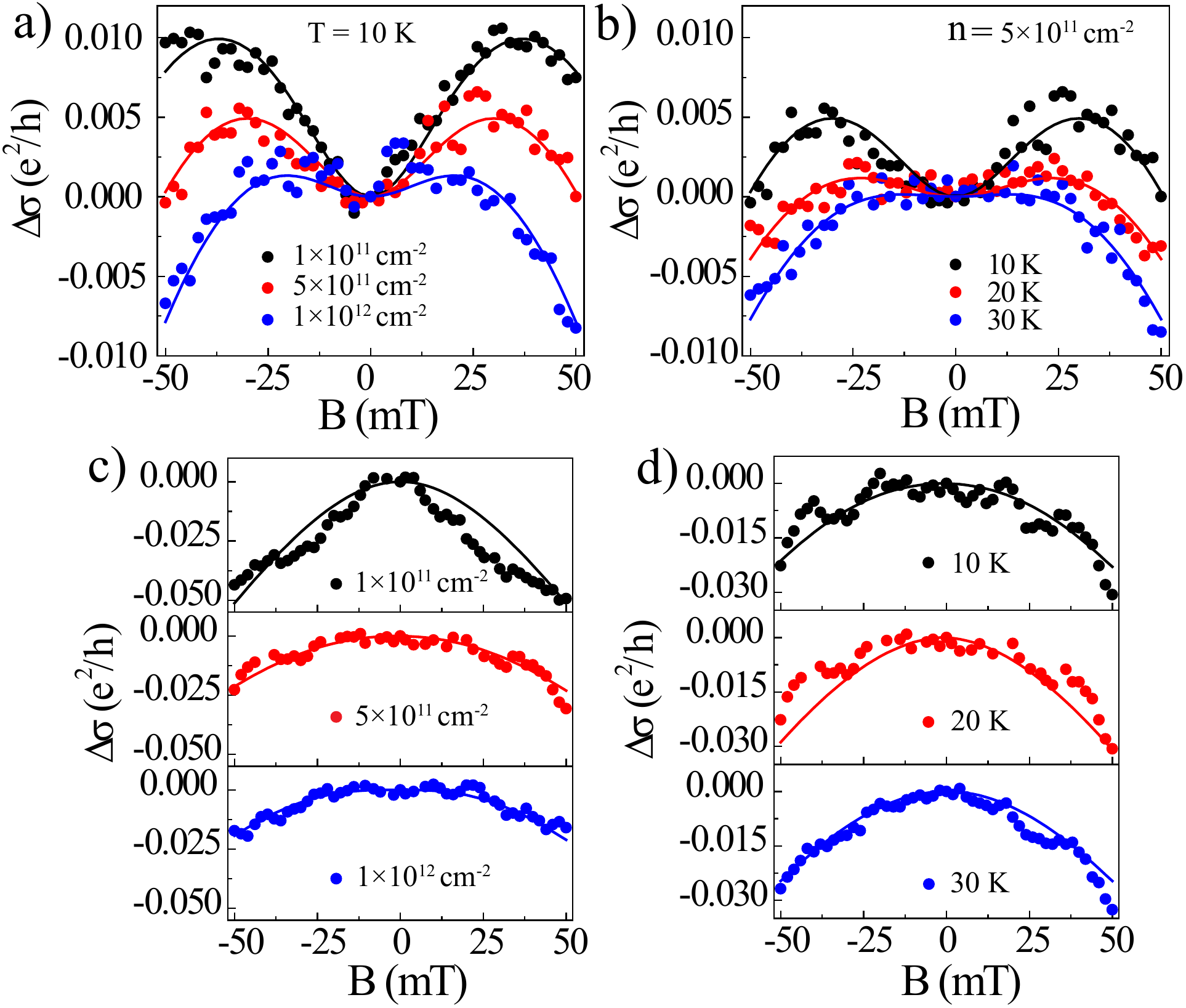}
		\caption{(a) The magneto-conductance data at different carrier concentrations at 10 K near CDC. (b) The magneto-conductance data at different temperatures for $n = 5\times10^{11} cm^{-2}$ near CDC. (c) The magneto-conductance data near PDC at different carrier concentrations for 10 K and (d) at different temperatures for $n = 5\times10^{11} cm^{-2}$. Solid curves are fits to Eq. (1)}
		\label{fig:image2}
	\end{figure}

It can be seen from Fig. 2 and 3 that WL becomes weaker with increasing temperature at CDC, which is expected due to the decrement of $\tau_{\phi}$ ($l_{\phi}$) with increasing temperature. However, it can be noticed that $\tau_{\phi}$ ($l_{\phi}$) dependence is much weaker at PDC reflecting a very weak dependence of WAL on temperature. With increasing density WAL at PDC becomes weaker because of the decrement of $\tau_{i}$ with increasing $n$\cite{tikhonenko2009transition}. However, at CDC with increasing density the WL becomes weaker, which can be understood by looking at density dependence of $\tau_{\phi}$ and $\tau_{i}$ (SI), where the dominance of $\tau_{\phi}$ over $\tau_{i}$ determines the trend.

As mentioned before the ratio of $\frac{\tau_{\phi}}{\tau_{i}}$ determines whether the MC will be WL or WAL. It can be seen from SI that  $\tau_{i}$ is one order smaller at CDC as compared to the values at PDC and as a consequence we observe the phase diagram in Fig. 3. The large variation of $\tau_{i}$ from PDC to CDC in the same device may be understood by considering the transition of Berry phase from PDC to CDC\cite{wang2015topological}. The $\pi$ Berry phase at PDC protect the inter-valley scattering from long range scatterers unless there are sharp defects in the device, resulting in higher value of $\tau_{i}$. However, at CDC in the new super-lattice Brillouin zone, the $K$ and $K^{'}$ valleys are connected by small momentum change. Thus, even the long range scatterers can contribute to inter valley scattering significantly since the back-scattering protection is lifted as a consequence of $2\pi$ Berry phase\cite{song2015topological}, giving rise to smaller value of $\tau_{i}$.

\begin{center}
	\begin{figure}[ht!]
		\includegraphics[width=.48\textwidth]{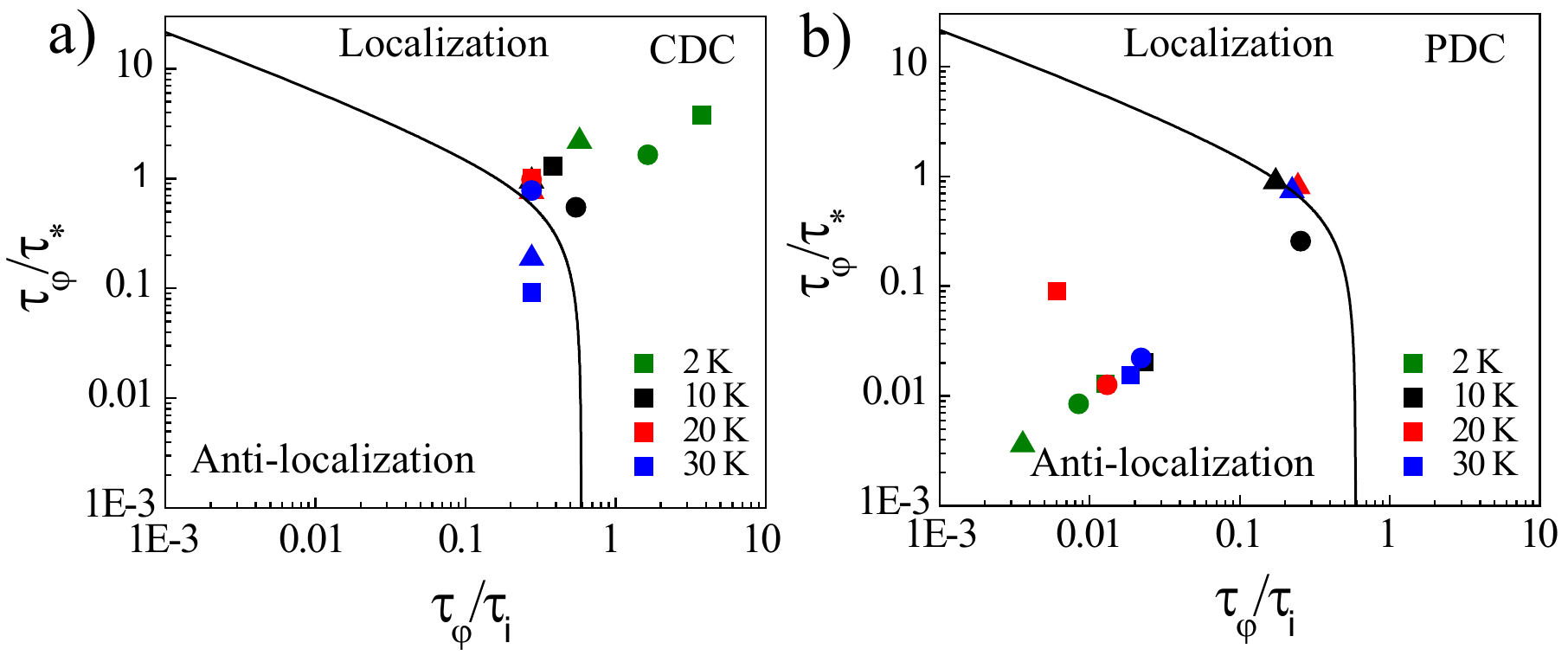}
		\caption{The different scattering times obtained by fitting the experimental data near CDC (a) and near PDC (b). The square, circle and triangular symbols represent carrier concentrations of $1\times10^{11} cm^{-2}, 5\times10^{11} cm^{-2}$ and $1\times10^{12} cm^{-2}$, respectively, where as different colors represent the data for different temperatures. The solid line separates the weak localization and the weak anti-localization region. It can be seen that majority of the points near CDC lie in weak localization region while points close to PDC lie in weak anti-localization region.}
		\label{fig:image3}
	\end{figure}
\end{center}

Although the change of Berry phase can explain the observed transition from WAL at PDC to WL at CDC by tuning Fermi energy in the same device, but the establishment of the one to one correspondence between the transition of localization and Berry phase change is essential to understand the origin of these properties. In order to determine the Berry phase experimentally we carry out quantum oscillation measurements.

Fig. 4 (a) shows the Landau Level (LL) spectrum (fan diagram) as a function of magnetic field and gate voltage at T = 2 K. From the conductance data the LLs originating from different Dirac points and their crossing are clearly visible. To calculate the Berry phase we follow the LLs maxima or minima originating from the PDC and how it changes after crossing the CDC. The vertical cut lines of Fig. 4 (a) at $V_{BG} = -30$ V and $V_{BG} = -55$ V are plotted in Fig. 4 (b) as a function of filling factor, $\nu = nh/4eB$, where the $n$ is determined from the PDC. Similar to Ref \cite{wang2015topological} we also observe a clear $\pi$ phase shift between the PDC and CDC as shown in Fig. 4 (b). Similar to Fig. 4(b), the phase shift can also be obtained from an analysis of fan diagram [Fig. 4 (a)] in terms of 1/B, where the modulation of the resistance can be written as \cite{zhang2005experimental,novoselov2005two}
$\Delta R_{xx} = R(B,T)cos [2\pi(B_{F}/B +1/2+\beta)]$ where $R(B,T)$ is the prefactor, $B_{F}$ is the frequency of oscillation in 1/B and $\beta$ is the associated Berry phase, in the range $0<\beta<1$. The $\beta$= 0.5 and 0 corresponds to Berry phase of $\pi$ and $2\pi$, respectively. We first locate the peaks and valleys of the oscillations in terms of 1/B along the vertical dashed lines in Fig. 4 (a) and then plot them against their Landau index N, which is shown in Fig. 4(c) for $V_{BG}$ = -30 V (blue) and $V_{BG}$ = -55 V (red). The slope of the linear fit gives the oscillation frequency related to carrier concentration where as the intercept yields the Berry phase, $\beta$ in units of $2\pi$. It can be clearly seen from the intercepts in Fig. 4 (c) that there is $\pi$ phase shift as we cross the CDC.

	\begin{figure}[ht!]
		\includegraphics[width=0.48\textwidth]{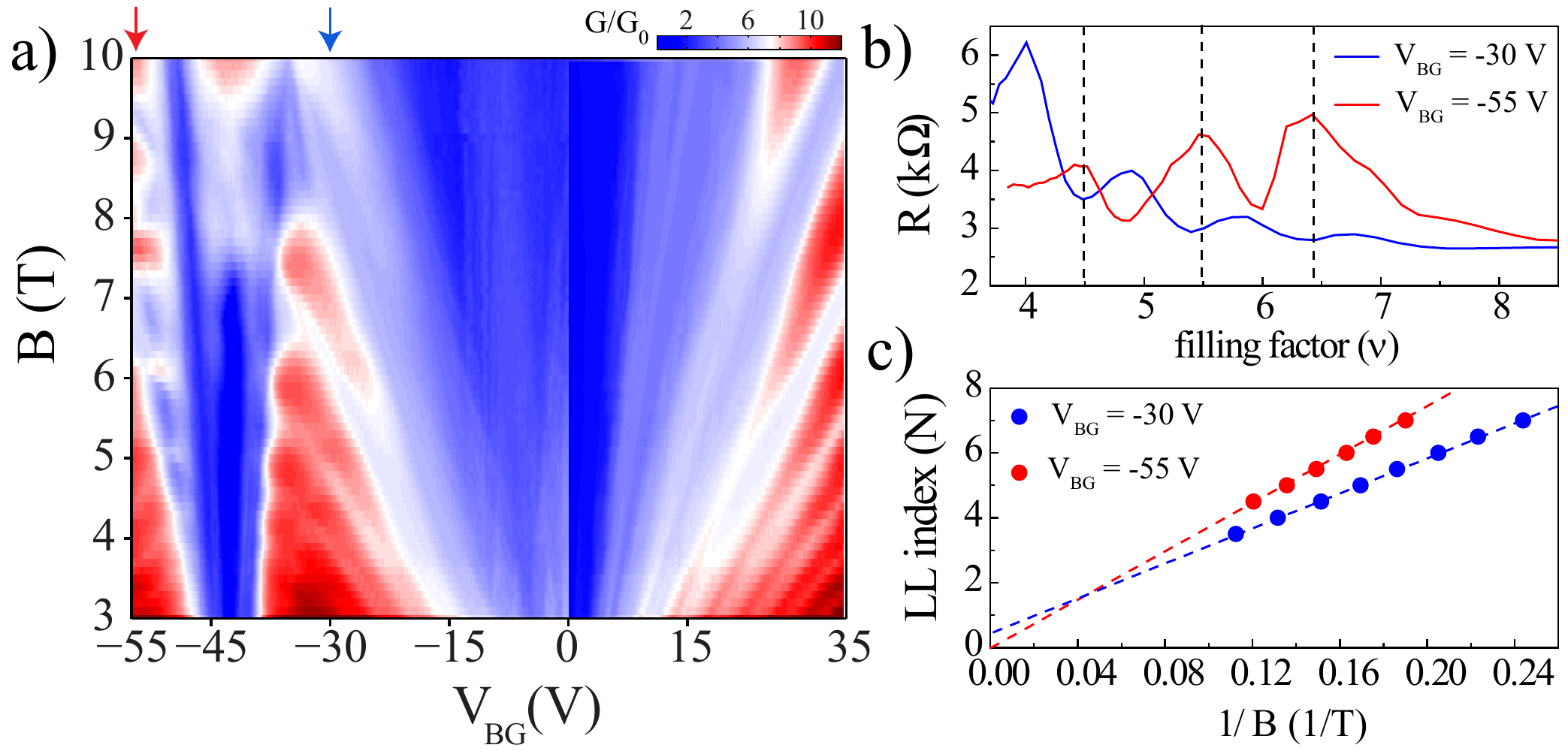}
		\caption{(a) The Landau fan diagram upto 10 T. Clear LL can be seen originating from PDC and CDC. (b) The resistance as a function of filling factor which are obtained from the vertical cut line of Fig. 4 (a) at two different gate voltages $V_{BG}=-30 V$ (blue line) and for $V_{BG}= -55 V$ (red line). It can be clearly seen that they are out of phase. (c) The 1/B value of the Nth minimum (Nth + 1/2 maximum) for the data shown in Fig 4 (b). Two different intercept can be seen on the y axis, 1/2 and 0, corresponding to Berry phase of $\pi$ and zero ($2\pi$) for PDC and CDC, respectively.}
		\label{fig:image4}
	\end{figure}

Next, we supplement these observations with detailed calculations of the Berry phase using realistic band structures. The low-energy electronic structure of the SLG/hBN setup can be modeled within the tight-binding approximation, with its parameters obtained from {\it ab-initio} calculations\cite{jung2014ab}. The resulting band structure shows cloning of the Dirac cone into six more Dirac cones around the $K$ and $K^{'}$ points. The computed dispersion for the PDC and CDC are schematically shown in the inset of Fig.~1(b), while the corresponding Fermi surface cuts at two representative energies determined by the experimental $n$ are shown in Fig. 5(c) and 5(d). While the properties of PDCs remain very much intact to their SLG counterpart, the CDCs show some interesting properties; some of which  are characteristically similar to the PDCs, while the others are different. One of the striking differences is that the CDCs possess a finite band gap, similar to BLG. Yet, unlike in BLG, where each PDC has zero Berry flux, here we show that different CDCs have finite but opposite Berry curvature, and the net Berry flux vanishes.

	\begin{figure}[ht!]
		\includegraphics[width=0.48\textwidth]{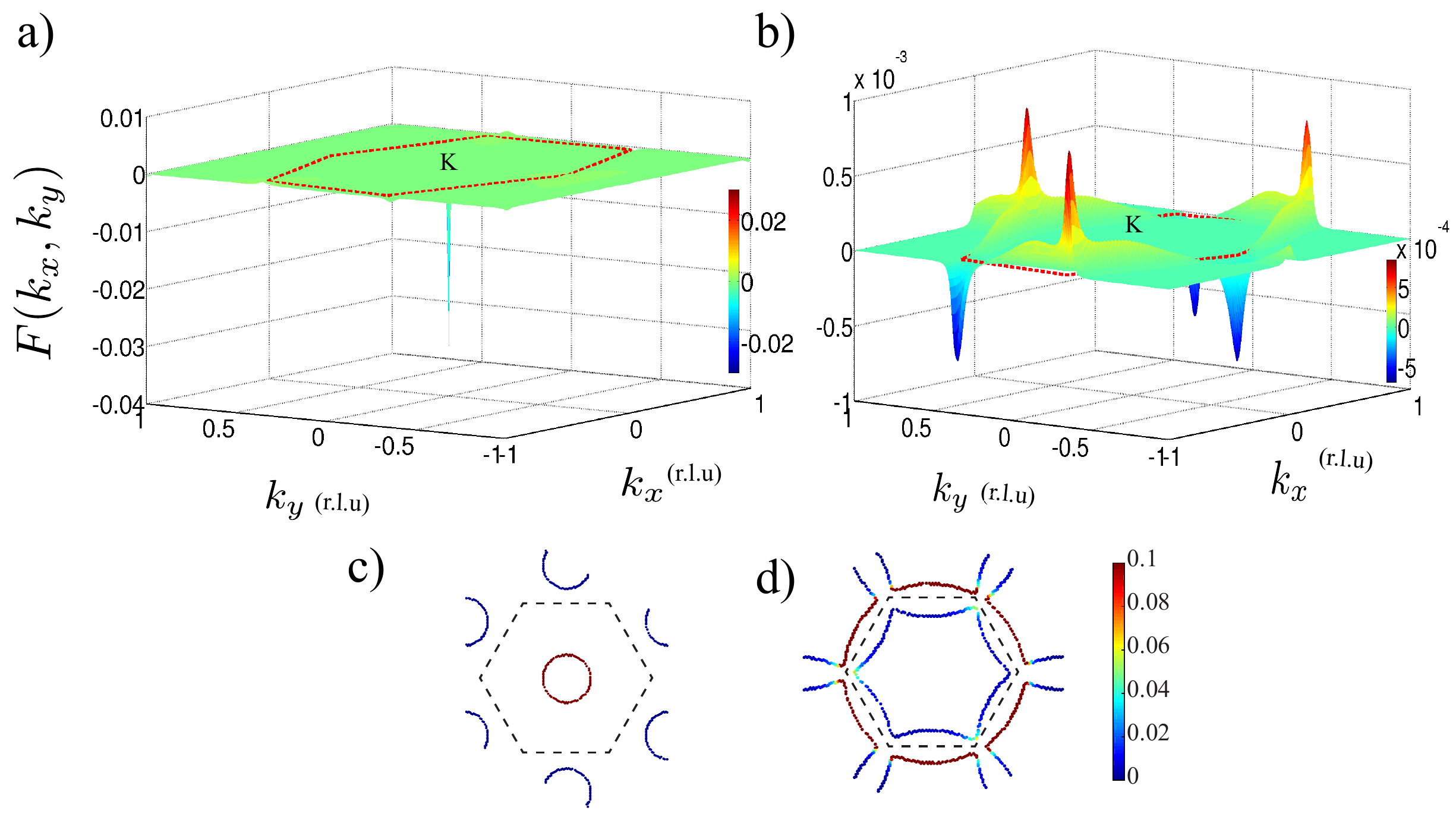}
		\caption{ Berry curvature variation within the moir$\acute{e}$ superlattice Brillouin zone (BZ)(red hexagon). $K$ indicates the Dirac point of SLG. (a) Strong Berry curvature peak at the superlattice BZ center near PDC. (b) Berry curvature around the CDC within superlattice BZ showing alternate positive and negative variation. (r.l.u = reciprocal lattice unit) (c) The Fermi surface cut at -100 meV from the PDC encloses only one $K$ point. (d) The Fermi surface cut at -40 meV from CDC. The Fermi surface in this reduced superlattice BZ is shared by all the six new $K$ points. The color bar indicates the weightage of the main band and shadow band.}
		\label{fig:image5}
	\end{figure}

The tight-binding model of moir$\acute{e}$ pattern is extensively discussed in the literature\cite{jung2014ab}, and we use the same model. We calculate Berry curvature, $F$ using the Kubo formula in the 2D momentum space (torus). In the context of graphene and topological insulator, it is known that the Berry curvature obtains a singular peak if there is a non-trivial band inversion at a single ${\bf K}$-point (or in a nodal ring) as shown in Fig. 5(a). In the presence of time-reversal the Berry curvature at $K$ and $K^{'}$ points of PDCs are exactly equal but opposite. As the chemical potential is tuned to the CDC, we notice that at each CDC, there occurs an additional band inversion (Fig.~1(b) inset). Here the main bands in the reduced moir$\acute{e}$ Brillouin zone (BZ) and the shadow bands from outside the BZ undergo inversion and produces additional Dirac cones. Although due to weak hBN potential, these cones have a finite gap, but owing to the associated band inversion, around the each gapped CDCs obtain finite Berry curvature. As expected, the Berry curvature at each CDCs are finite but opposite in sign as shown in Fig. 5(b). The net Berry flux or Berry phase of a band can be defined as 
\begin{eqnarray}
\gamma_{n}=\int_{\Omega_{\rm BZ}} dk_{x} dk_{y} F^{n}({\bf k}),
\end{eqnarray} 
where $n$ is the band index, and $\Omega_{\rm BZ}$ is the phase space area of the BZ.

The Berry phase will be zero or $2\pi$ if the Fermi surface encloses all the 6 $K$ points around the superlattice BZ. From our quantum oscillation measurement (period of $1/B$) we could verify this by measuring the area of the Fermi surface around PDC and after crossing the CDC (SI). It turns out that indeed at PDC Fermi surface is confined to only one of the $K$ points whereas after crossing the CDC the Fermi surface captures entire superlattice BZ. This has been also verified theoretically as shown in Fig. 5(c) and 5(d), where the Fermi surface cuts are shown at $\sim$ 100 meV below the PDC and $\sim$ 40 meV below the CDC, respectively. It can be seen that after crossing the CDC the Fermi surface originating from the shadow bands encloses none of the six $K$ points completely in the reduced BZ, rather partially captures all the six $K$ points, which is consistent with the literature \cite{song2015topological}. Although the Fermi surfaces very close to the CDC, which can enclose only one CDC point, are within $\sim$ 25 meV from the CDC (SI) and thus remains technically challenging to carry out the localization study very close to the CDC due to the presence of electron-hole puddles and charge inhomogeneities\cite{ki2008inelastic,tikhonenko2008weak,moser2010magnetotransport,iqbal2014enhanced,chandni2015transport}.

In conclusion, our magneto conductance data shows a clear transition from WAL at PDC to WL at CDC in a single device by tuning the Fermi energy. We also observe that the quantum oscillations data shows the shift of Berry phase from  $\pi$ at PDC to 2$\pi$ at CDC. These two experimental observations with theoretical support strongly suggest the effect of Berry phase on localization physics, which will help to understand the quantum corrections to the conductivity in multiple Dirac cones in graphene superlattice as well as will stimulate the further theoretical studies on localization physics in graphene superlattice.

\pagebreak
\begin{center}
	\textbf{\large Supplementary Material: Localization physics in graphene Moire superlattices}
\end{center}
\begin{center}
	Chandan Kumar${}^{1}$, Saurabh Kumar Srivastav${}^{1}$, Priyo Adhikary${}^{1}$, Sumilan Banerjee${}^{1}$, Tanmoy Das${}^{1}$ and Anindya Das${}^1$
	
	\it{${}^1$Department of Physics, Indian Institute of Science, Bangalore 560012, India}
\end{center}	
	
	\section{Magneto-conductance data}
	In the main text, data is presented at 2 K and 10 K near CDC and at 10 K near PDC. Here, Fig. 6 and Fig. 7 shows the density dependence of magneto conductance near CDC and PDC, respectively at 20 K and 30 K .
	\begin{figure}[h]
		\includegraphics[width=0.8\textwidth]{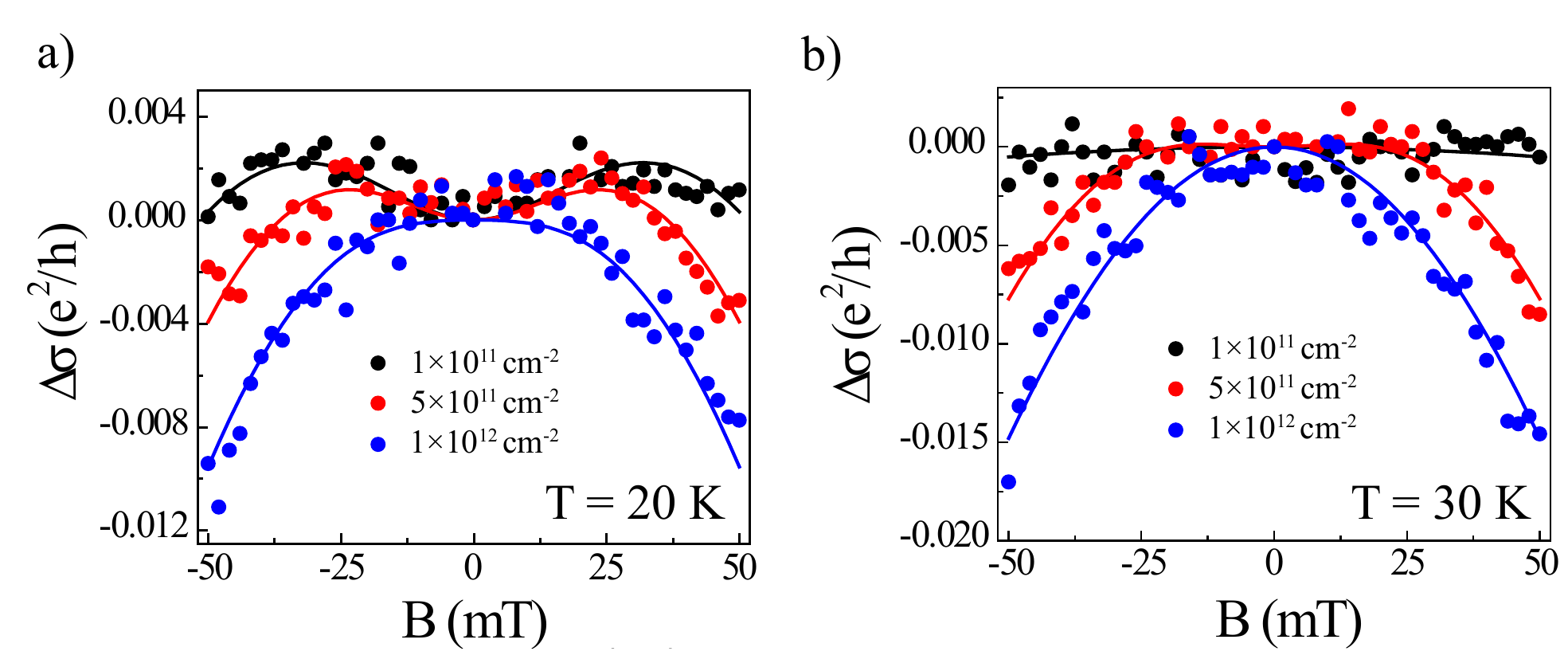}
		\caption{(Color Online) Magneto-conductance data near CDC at 20 K (a) and 30 K (b) at different densities.}
	\end{figure}
	
	\begin{figure}[h]
		\includegraphics[width=0.6\textwidth]{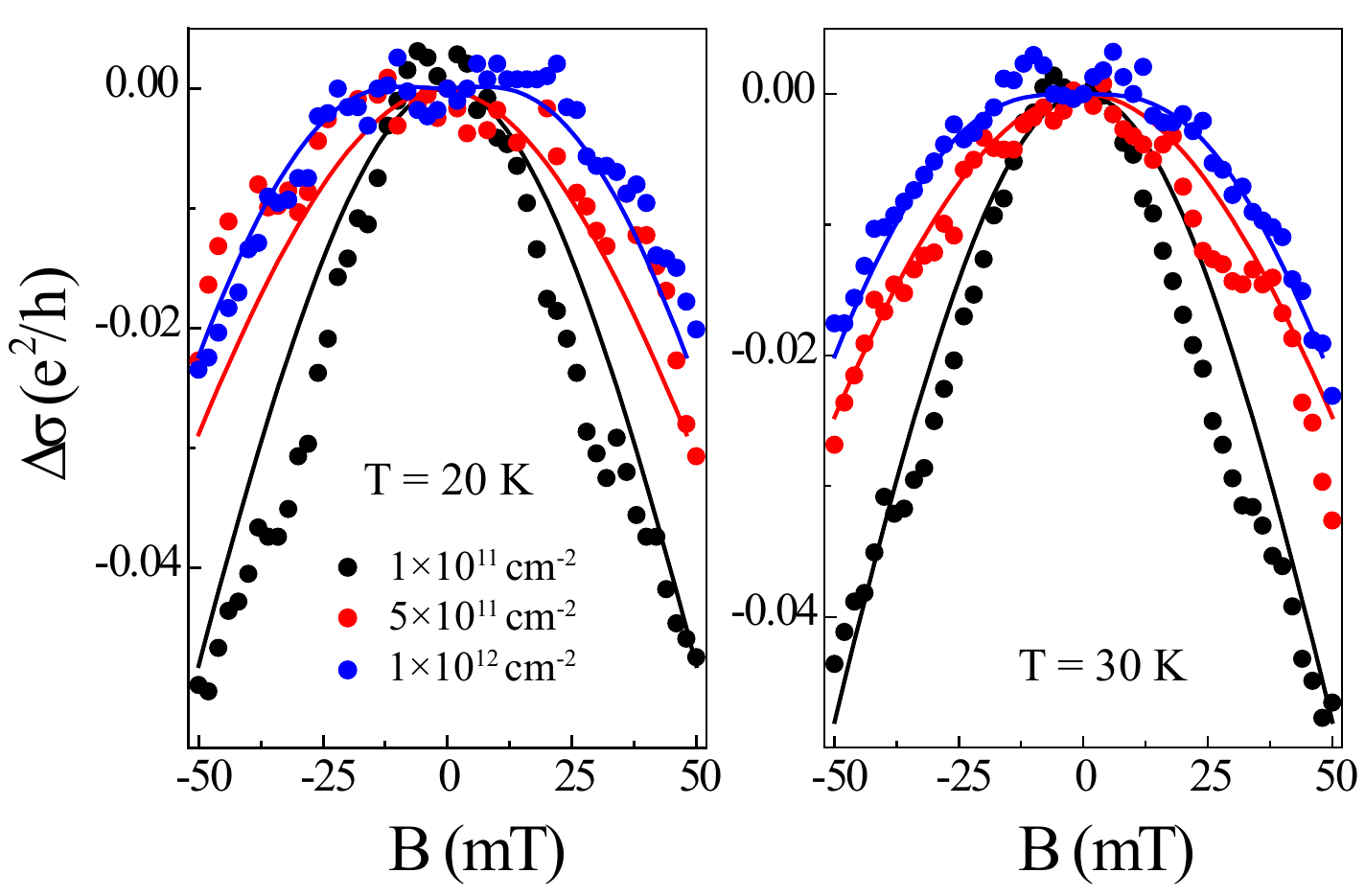}
		\caption{(Color Online) Magneto-conductance data near PDC at 20 K (a) and 30 K (b) at different densities.}
	\end{figure}
	\newpage	
	\section{CALCULATION OF DIFFERENT SCATTERING RATES}
	
	The magneto-conductance data of Fig. 1d and Fig. 2 of the main text in manuscript and supplementary Fig. 6 and 7 are fitted with following equation\cite{mccann2006weak}
	\begin{equation}
	\Delta \sigma (B) = \frac{e^2}{\pi h} \left[ F\left(\frac{B}{B_{\phi}}\right)-F\left(\frac{B}{B_{\phi}+2B_{i}}\right)-2F\left(\frac{B}{B_{\phi}+B_{i}+B_{*}}\right) \right]
	\end{equation}
	where $F (z) = ln(z)+\psi \left( \frac{1}{2} +\frac{1}{z}\right)$ , $B_{\phi,i,*}=\frac{\hbar}{4De}\tau_{\phi,i,*}^{-1}$\\ 
	Here $\psi(x)$ is the digamma function, $\tau_{\phi}^{-1}$ is the phase breaking rate, D is the diffusion coefficient, $\tau_{i}^{-1}$ is the intervalley scattering rate and  $\tau_{*}^{-1}$ is the intravalley scattering rate.
	The diffusion coefficient is calculated using the expression\cite{tikhonenko2008weak,ki2008inelastic}: 
	$$ D= \frac{1}{2}v_Fl_m=\frac{hv_F\sigma}{2e^2k_F}$$
	where $l_m$ is the mean free path and $k_F= \sqrt{\pi n}$.\\
	
	The momentum relaxation rate ($\tau_p^{-1}$) is obtained using relation $\tau_p^{-1}=\frac{v_F}{l_m}$ and we found it to be  highest in the system. The different scattering rates obtained from fitting are listed in the table below.
	\\
	\\
	\begin{tabular}{ |p{3cm}||p{3cm}|p{3cm}|p{3cm}||p{3cm}|  }
		\hline
		\multicolumn{5}{|c|}{Primary Dirac Cone} \\
		\hline
		Density ($ cm^{-2}$) & $\tau_p^{-1}$ ($sec^{-1}$) &$\tau_{\phi}^{-1}$ ($sec^{-1}$)&$\tau_{i}^{-1}$ ($sec^{-1}$)&$\tau_{*}^{-1}$ ($sec^{-1}$)\\
		\hline
		$1 \times 10^{11}$ & $ \sim 5.2 \times 10^{12}$ &$ \sim 2.6 \times 10^{11}$& $ \sim 3.3 \times 10^{9}$ & $ \sim 3.2 \times 10^{9}$\\
		$5 \times 10^{11}$ & $ \sim 6.0 \times 10^{12}$ &$\sim 8.8 \times 10^{11}$& $ \sim 7.4 \times 10^{9}$ & $ \sim 7.5 \times 10^{9}$\\
		$1 \times 10^{12}$ & $ \sim 8 \times 10^{13}$ &$\sim 3 \times 10^{12}$& $ \sim 1 \times 10^{10}$ & $ \sim 1.2 \times 10^{10}$\\
		\hline
	\end{tabular}\\
	\begin{tabular}{ |p{3cm}||p{3cm}|p{3cm}|p{3cm}||p{3cm}|  }
		\hline
		\multicolumn{5}{|c|}{Cloned Dirac Cone} \\
		\hline	
		Density ($ cm^{-2}$) & $\tau_p^{-1}$ ($sec^{-1}$) &$\tau_{\phi}^{-1}$ ($sec^{-1}$)&$\tau_{i}^{-1}$ ($sec^{-1}$)&$\tau_{*}^{-1}$ ($sec^{-1}$)\\
		\hline
		$1 \times 10^{11}$ & $ \sim 8 \times 10^{12}$&$ \sim 2.1 \times 10^{10}$&  $ \sim 1 \times 10^{11}$ & $ \sim 9.6 \times 10^{10}$\\
		$5 \times 10^{11}$ & $ \sim 7.6 \times 10^{12}$&$ \sim 2 \times 10^{11}$ & $ \sim 4.6 \times 10^{11}$ & $ \sim 4.3 \times 10^{11}$\\
		$1 \times 10^{12}$ & $\sim 8 \times 10^{13}$ &$ \sim 2.9 \times 10^{11}$&  $ \sim 4.0 \times 10^{11}$ & $ \sim 1\times 10^{12}$\\
		\hline
	\end{tabular}
	\\
	\\
	\\
	Note that in calculating the different scattering rates diffusion coefficients must be taken into account but the diffusion coefficients do not play any role in Fig. 3 of the main manuscript, where the ratio of different scattering rates are plotted.\\
	
	Equation (4) can be rewritten as:\cite{tikhonenko2008weak}
	\begin{equation} 
	\Delta \sigma (B) = \frac{e^2}{\pi h} \left[ F\left(\frac{\tau_B^{-1}}{\tau_\phi^{-1}}\right)-F\left(\frac{\tau_B^{-1}}{\tau_\phi^{-1}+2\tau_i^{-1}}\right)-2F\left(\frac{\tau_B^{-1}}{\tau_\phi^{-1}+\tau_i^{-1}+\tau_*^{-1}}\right) \right]
	\end{equation}
	The equation (5) is mentioned in the main text of the manuscript.\\
	\newpage
	\section{DEPHASING MECHANISM}
	In graphene electron-electron interaction through Nyquist scattering is considered to be the major source of dephasing mechanism\cite{altshuler1982effects,tikhonenko2009transition}. $L_{\phi}$ in diffusive regime is defined as\cite{tikhonenko2009transition}:
	\begin{equation} 
	L_{\phi}^{-2}=\alpha \frac{K_B T}{D} \frac{lng}{g}
	\end{equation}
	where $g=\sigma/(e^2/h)$. Diffusive regime corresponds to $K_B T \tau_p/\hbar <1$ i.e an electron undergoes many collisions with defects during the interaction time $\hbar/(K_B T)$.
	For our sample we find that $K_B T \tau_p/\hbar$ varies from $0.036 $ to $0.545$ in the temperature range 2 K to 30 K.
	In Fig. 8 we plot $\L_{\phi}$ (extracted from the fitting of the magneto conductance data near PDC and CDC) as a function of temperature. The solid line is the $1/\sqrt{T}$ fit at  PDC and CDC. We see that $L_{\phi}$  varies approximately as $1/\sqrt{T}$ for both PDC and CDC. This implies that the electron-electron interaction is the major source of dephasing mechanism.\\
	Using $\sigma = \frac{2e^2}{h} K_F l$, Eq. (6) can be rewritten as
	\begin{equation} 
	\tau_{\phi}^{-1}= \alpha \frac{K_B T}{2E_F \tau_p} ln(\frac{2E_F \tau_p}{\hbar})
	\end{equation}
	We find that  $\tau_{\phi}^{-1}$ is directly proportional to temperature.\\
	It is also observed that in graphene the electron-phonon scattering $\tau_{e-ph}^{-1}$ is also directly proportional to temperature and is given as: \cite{stauber2007electronic,hwang2008acoustic,tikhonenko2009transition}
	\begin{equation} 
	\tau_{e-ph}^{-1}= \frac{1}{\hbar^3}\frac{E_F}{4V_F^2}\frac{D_a^2}{\rho_m V_{ph}^2} K_B T
	\end{equation}
	
	where $E_F$ is the Fermi energy, $V_F$ is the Fermi velocity, $V_{ph}$ is the speed of sound, $\rho_{m}$ is the density of graphene and $D_{a}$ is the deformation potential. Thus to unambiguously prove that the electron-electron is the main source of dephasing mechanism, we plot $\tau_{\phi}^{-1}$ (with scatter points) at CDC and $\tau_{e-ph}^{-1}$ with dashed line in the inset of Fig. 8. The electron-phonon scattering rate is obtained using the parameters $\rho_{m} = 7.6\times 10^{-7} Kg m^{-2}$, $V_F = 10^6 ms^{-1}$, $V_{ph}= 2\times 10^4 ms^{-1}$ and $D_{a}= 18 eV$ \cite{tikhonenko2009transition}. We find  that electron-phonon scattering rate is too small to explain our data. The solid line is a linear fit to $\tau_{\phi}^{-1}$ at CDC. 
	
	\begin{figure}[h]
		\includegraphics[width=0.5\textwidth]{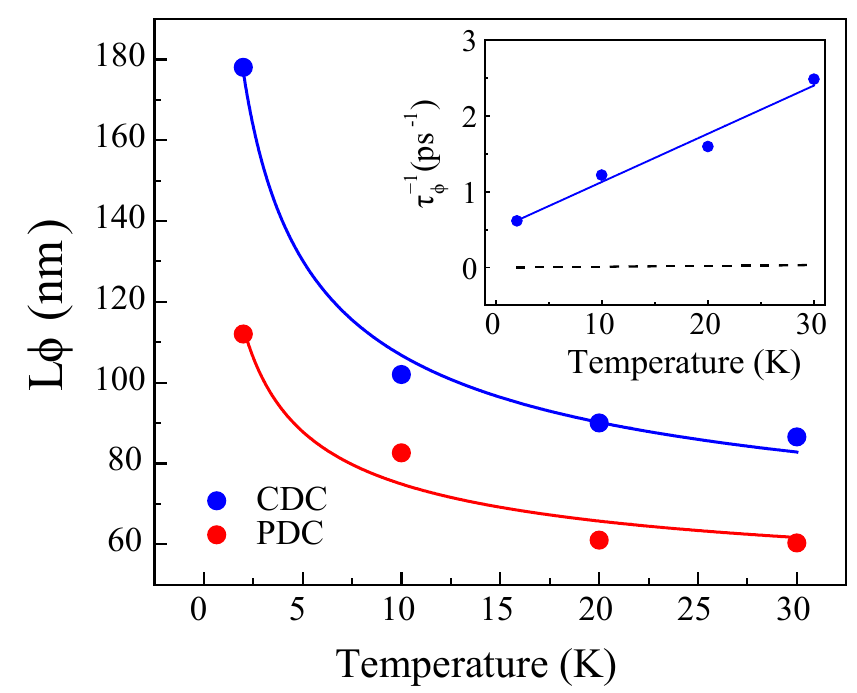}
		\caption{(Color Online) The variation of $L_{\phi}$ as a function of temperature at $n = 5 \times 10^{10} cm^{-2}$ at CDC and PDC. The solid line is a $1/\sqrt{T}$ to the data points. Inset: Temperature dependence of $\tau_{\phi}^{-1}$. The solid line is fit to electron- electron scattering rate. The dashed line is the electron-phonon rate calculated using Eq. (8).}
	\end{figure}
	\newpage
	
	\section{Fermi surface calculation}
	Fig. 9 shows the trans-conductance ($dG/dV_{BG}$) as a function of magnetic field and the back gate voltage at 240 mK. Clear Landau levels can be seen originating from PDC and CDC. The dashed line shows the landau levels originating from the PDC and the vertical solid lines are drawn at -21V, -29 V and -55 V. \\
	The oscillation period of Landau levels in $1/B$ at a given gate voltage is calculated by locating the Landau levels maxima (denoted by the yellow circles) and plotting them against the respective $1/B$ value. 
	The Fermi surface area (S) is calculated using following relation:
	\begin{equation} 
	S= \frac{2\pi e}{\hbar c \Delta (1/B)}
	\end{equation}
	From this area we can estimate the density using the following equation:
	\begin{equation} 
	S= \pi {k_{F}}^{2}  = n {\pi}^{2}
	\end{equation}
	The table below summaries the calculation of density and Fermi surface at different gate voltages.
	
	\begin{figure}[h]
		\includegraphics[width=0.5\textwidth]{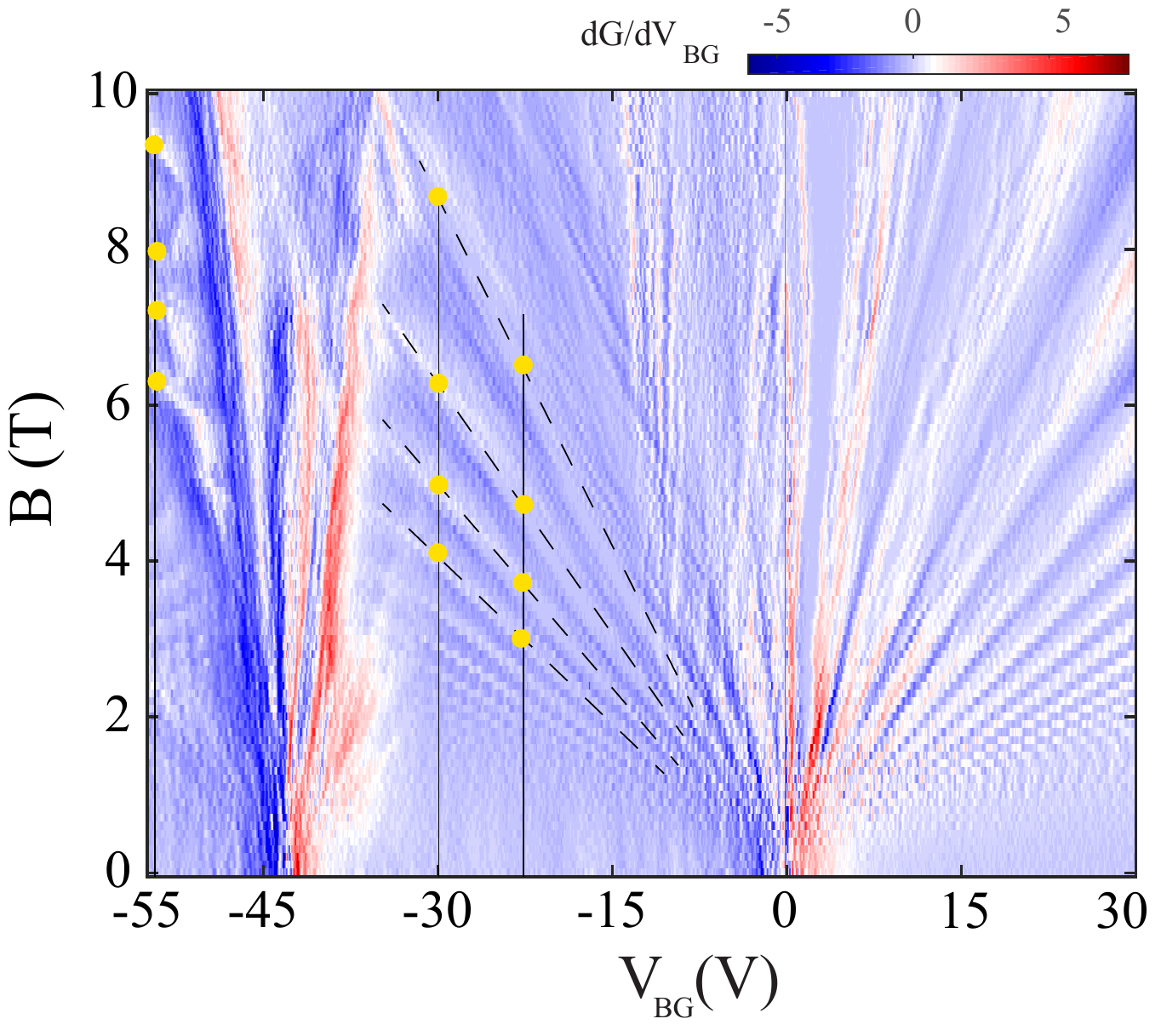}
		\caption{(Color Online) Trans-conductance ($dG/dV_{BG}$) as a function of magnetic field and back gate voltage at zero top gate voltage. The dashed line shows the landau levels originating from the PDC. The vertical solid line is drawn at -21 V, -29 V and -55 V. The yellow filled circles denote the Landau level maxima at -21 V, -29 V and -55 V.}
	\end{figure}
	\begin{tabular}{ |p{3cm}||p{3cm}|p{3cm}|p{3cm}||p{3cm}|  }
		\hline
		\multicolumn{5}{|c|}{Density and Fermi surface calculation} \\
		\hline
		Gate Voltage (V)& Density calculation using $n= \frac{C_{bg}}{e} (V-V_{PDC}) $ ($ cm^{-2}$)& Oscillation Period in 1/B ($T^{-1}$) & Fermi surface,S ($cm^{-2}$) & Density calculation from S ($cm^{-2}$) \\
		\hline
		$-21$ & $ \sim 1.5\times 10^{12}$& 0.0615  & $1.56 \times 10^{13} $& $\sim 1.57\times 10^{12} $ \\
		\hline
		$-29$ & $ \sim 2.1\times10^{12} $  & 0.0463 & $2.1 \times 10^{13}  $& $\sim 2.08 \times10^{12} $  \\
		\hline
		$-55$ & $ \sim 4.0\times10^{12} $  & 0.0232& $4.1 \times 10^{13}    $&  $\sim 4.16 \times10^{12}$  \\
		\hline
	\end{tabular}\\
	\\
	It can be seen that the density value obtained using $ n= \frac{C_{bg}}{e} (V-V_{PDC})$ and Fermi surface are of very similar in magnitude.
	\newpage
	\section{Density of state and Fermi surface}
	The energy separation between the PDC and CDC is $ \sim190$ meV which corresponds to moire wavelength of $\sim 12$ nm and the angle mismatch of $\sim 1^{0}$ degree.
	\begin{figure}[h]
		\includegraphics[width=0.7\textwidth]{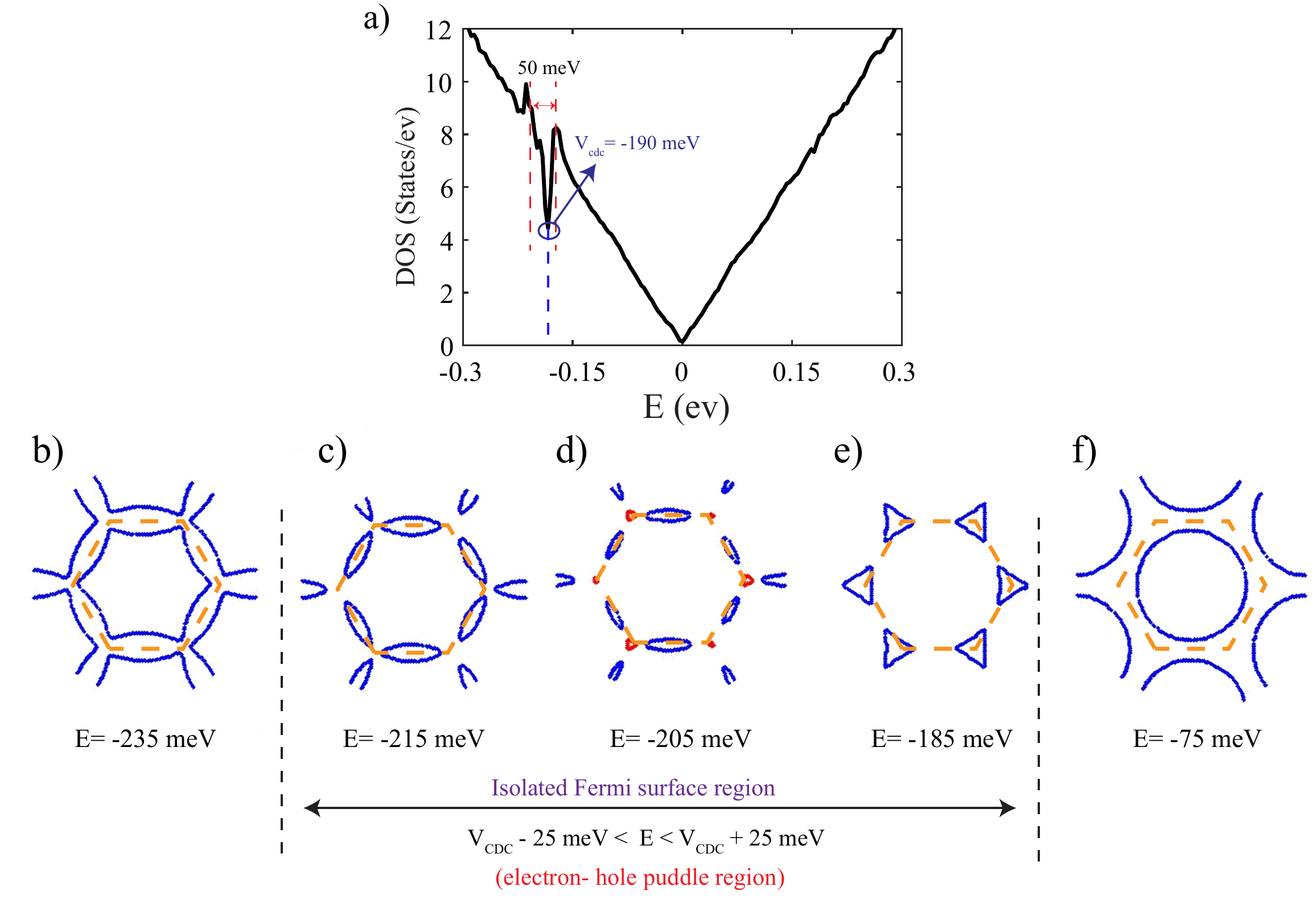}
		\caption{(Color Online) (a) Density of state calculated using tight binding model (b-f) Fermi surface plot at different energy.}
	\end{figure}
	
	Figure 10 (a) shows the density of state (DOS) for our device parameter. We observe a dip in DOS in the hole doped region as compared to electron doped region, which is consistent with literature\cite{dean2013hofstadter,woods2014commensurate,yu2014hierarchy}. 
	Fig. 10 (b-f) shows the evolution of Fermi surface at different energy close to CDC. It is interesting to note that Fermi surface in a very small energy range of $\pm 25$ meV around CDC encloses only one of the Dirac cone around the CDC as shown in Fig. 10 (c-e). However, below -25 meV from CDC the Fermi surface in the new superlattice brillouin zone (BZ) is shared by all the  new six $K$ points as shown in Fig. 10b. We also calculate the charge in-homogeneity near CDC, $\delta n \sim 3 \times 10^{10} cm^{-2}$(will be discussed in the next section). This corresponds to Fermi energy broadening of $\sim 20$ meV. As a result we could study the localization physics in the energy range below -25meV from the CDC to avoid the effect of charge puddles on localization as mentioned in the literature\cite{ki2008inelastic,tikhonenko2008weak,moser2010magnetotransport,iqbal2014enhanced,chandni2015transport}.
	
	
	\section{Device Fabrication}
	
	The graphene super-lattice heterostructure is prepared by making stack of hBN/Gr/hBN using pick up technique\cite{zomer2011transfer} with the following steps: First a glass slide is prepared with a layer of PDMS and PPC. This glass slide is used to pick up hBN and graphene flakes, which were exfoliated separately on different $SiO_2$ substrate. The glass slide containing hBN/Gr stack is then aligned and transferred on another hBN which was exfoliated on a new $SiO_2$ substrate. The finally prepared stack of hBN/Gr/hBN is then cleaned in chloroform, acetone and IPA. This is followed by ion etching in $CHF_3$ and $O_2$ environment and Cr/Au evaporation at a base pressure of 1E-7 mbar. 
	With this technique the graphene remains pristine as it is not exposed to any environmental residue or pmma.\\
	\newpage
	\section{MOBILITY AND $\mathbf{\delta n}$ (in-homogeneity) EXTRACTION}
	
	Figure 11 (a) and (b) shows resistance as a function of back gate voltage ($V_{BG}$) at primary Dirac cone (PDC) and cloned Dirac cone (CDC), respectively. The mobility and the charge in-homogeneity are extracted by fitting the $R-V_{BG}$ curve, of figure 11 (a) and 11(b) with the following equation:\cite{venugopal2011effective}
	\begin{equation}
	R=R_c+\frac{l}{we\mu\sqrt{\delta n^2+n_{ch}^{2}}}
	\end{equation}
	where $R_c$ is the contact resistance, $l$ is the channel length, $w$ is the width, $\delta n$ is the charge in-homogeneity and $n_{ch}$ is the induced charge due to applied back gate voltage, $V_{BG}$.  The length and width of our device are $l= 1$ $\mu m$ and $w= 3.4$ $\mu m$, respectively. From the fitting, we obtain mobility of $\sim$ 22000 and 20000 $cm^2V^{-1}s^{-1}$ at PDC and CDC, respectively and charge in-homogeneity, $\delta n \sim 3 \times 10^{10} cm^{-2}$ .
	\begin{figure*}[ht!]
		\includegraphics[width=0.8\textwidth]{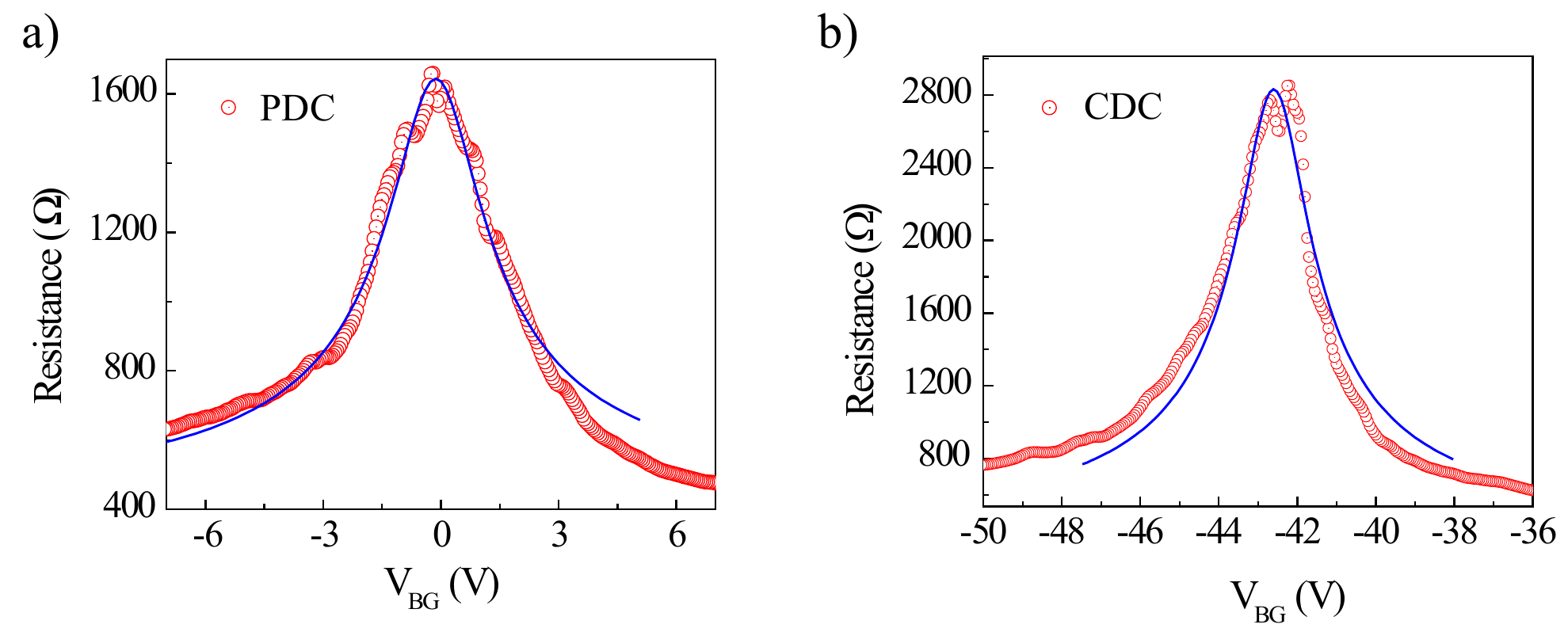}
		\caption{(Color Online) Resistance as a function of $V_{BG}$ for (a) PDC and (b) CDC at 2 K. The blue solid lines are the fitted lines with the equation (11)}
	\end{figure*}		
	
	The charge in-homogeneity is also calculated by plotting log of conductance Vs log of density ($n$). Figure 12 (a) and (b) shows log of conductance as a function of log $n$ for positive $V_{BG}$ near PDC and CDC, respectively. From the linear fit in Fig. 12 (a) and Fig. 12 (b), we obtain charge inhomogeneity of $\delta n \sim 3\times 10^{10} cm^{-2}$ .
	\begin{figure*}[ht!]
		\includegraphics[width=0.8\textwidth]{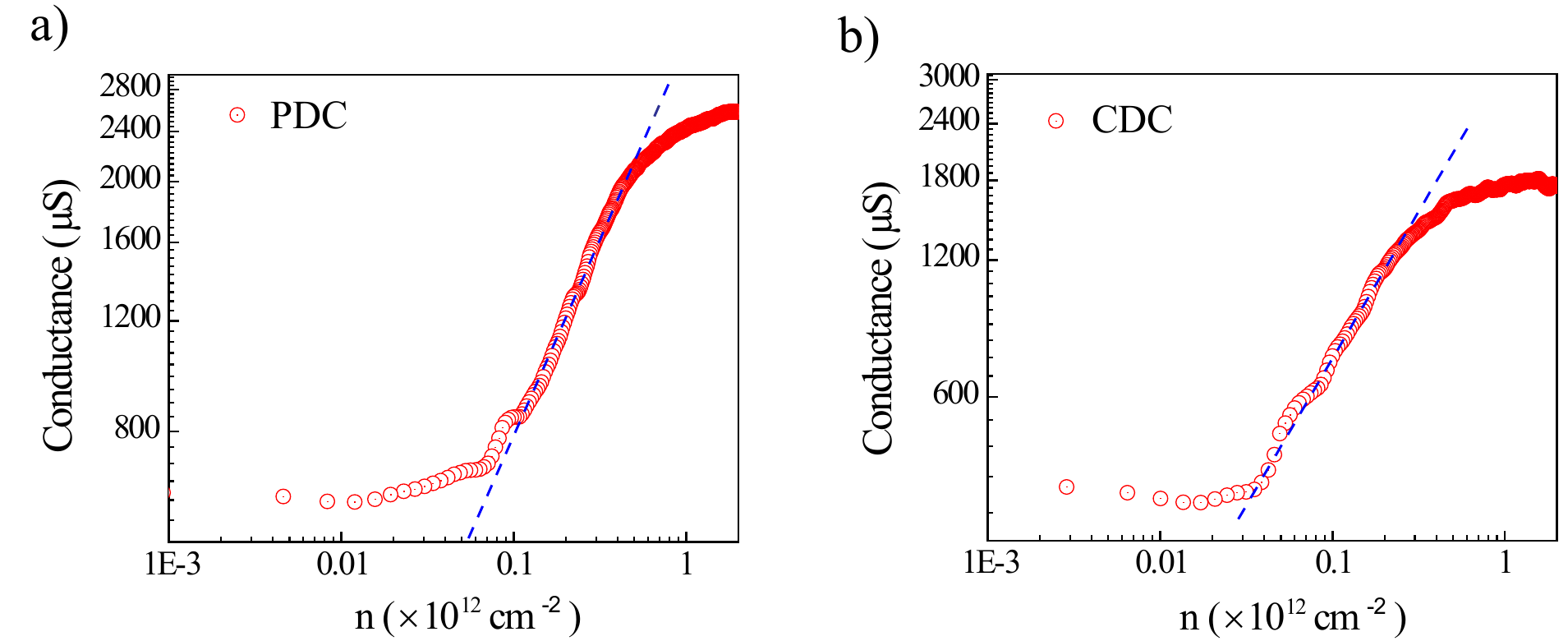}
		\caption{(Color Online) Conductance as a function of positive $V_{BG}$ in the log-log scale for (a) PDC and (b)CDC. The blue lines are the linear fits to the data to extract the charge in-homogeneities.}
	\end{figure*} 
	\newpage
	\section{BAND GAP AT CLONED DIRAC CONE}
	
	Figure 13 shows the $R-V_{BG}$ for different temperatures. It can be seen from the Fig. 13 that we do not observe any signature of significant band gap opening near CDC and PDC in our device. Although, our tight binding calculation predicts a small band gap opening near CDC but may be due to finite Fermi energy broadening the band gap opening is not observed in the experimental data. This is consistent with the earlier reports\cite{dean2013hofstadter,woods2014commensurate,yu2014hierarchy} where no gap was observed when the lattice mismatch between hBN and graphene is small($\sim 1^{0}$ degree).
	\begin{figure}[h]
		\includegraphics[width=0.5\textwidth]{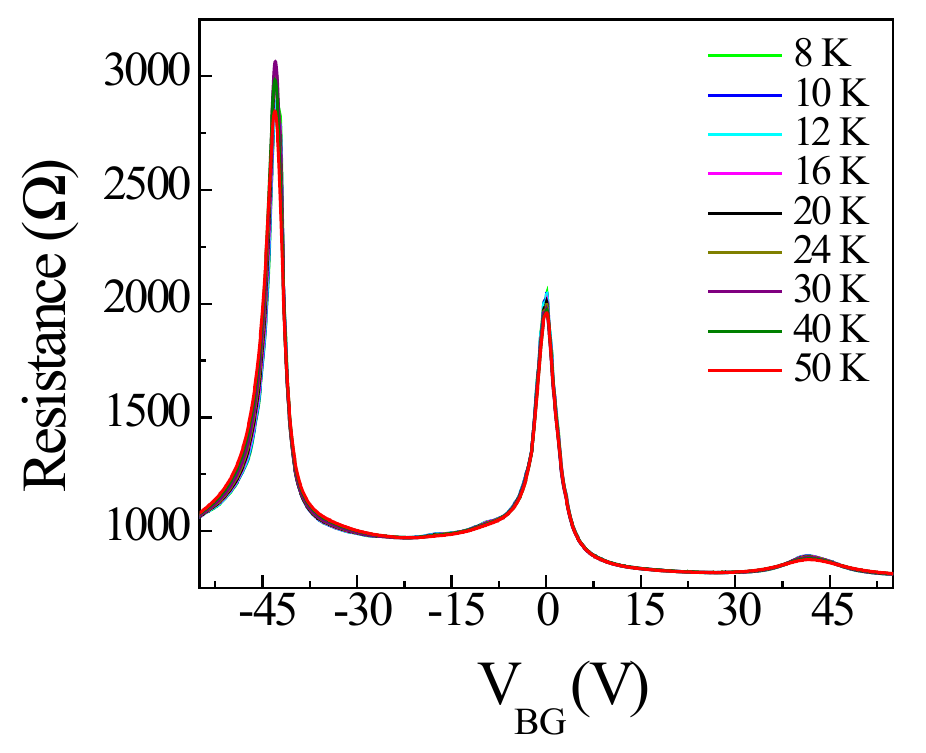}
		\caption{(Color Online) Resistance as a function of $V_{BG}$ for different temperatures.}
	\end{figure}
\bibliographystyle{apsrev4-1}
\bibliography{ref}

\begin{thebibliography}{58}%
\makeatletter
\providecommand \@ifxundefined [1]{%
 \@ifx{#1\undefined}
}%
\providecommand \@ifnum [1]{%
 \ifnum #1\expandafter \@firstoftwo
 \else \expandafter \@secondoftwo
 \fi
}%
\providecommand \@ifx [1]{%
 \ifx #1\expandafter \@firstoftwo
 \else \expandafter \@secondoftwo
 \fi
}%
\providecommand \natexlab [1]{#1}%
\providecommand \enquote  [1]{``#1''}%
\providecommand \bibnamefont  [1]{#1}%
\providecommand \bibfnamefont [1]{#1}%
\providecommand \citenamefont [1]{#1}%
\providecommand \href@noop [0]{\@secondoftwo}%
\providecommand \href [0]{\begingroup \@sanitize@url \@href}%
\providecommand \@href[1]{\@@startlink{#1}\@@href}%
\providecommand \@@href[1]{\endgroup#1\@@endlink}%
\providecommand \@sanitize@url [0]{\catcode `\\12\catcode `\$12\catcode
  `\&12\catcode `\#12\catcode `\^12\catcode `\_12\catcode `\%12\relax}%
\providecommand \@@startlink[1]{}%
\providecommand \@@endlink[0]{}%
\providecommand \url  [0]{\begingroup\@sanitize@url \@url }%
\providecommand \@url [1]{\endgroup\@href {#1}{\urlprefix }}%
\providecommand \urlprefix  [0]{URL }%
\providecommand \Eprint [0]{\href }%
\providecommand \doibase [0]{http://dx.doi.org/}%
\providecommand \selectlanguage [0]{\@gobble}%
\providecommand \bibinfo  [0]{\@secondoftwo}%
\providecommand \bibfield  [0]{\@secondoftwo}%
\providecommand \translation [1]{[#1]}%
\providecommand \BibitemOpen [0]{}%
\providecommand \bibitemStop [0]{}%
\providecommand \bibitemNoStop [0]{.\EOS\space}%
\providecommand \EOS [0]{\spacefactor3000\relax}%
\providecommand \BibitemShut  [1]{\csname bibitem#1\endcsname}%
\let\auto@bib@innerbib\@empty
\bibitem [{\citenamefont {Altshuler}\ \emph {et~al.}(1980)\citenamefont
  {Altshuler}, \citenamefont {Khmel'Nitzkii}, \citenamefont {Larkin},\ and\
  \citenamefont {Lee}}]{altshuler1980magnetoresistance}%
  \BibitemOpen
  \bibfield  {author} {\bibinfo {author} {\bibfnamefont {B.}~\bibnamefont
  {Altshuler}}, \bibinfo {author} {\bibfnamefont {D.}~\bibnamefont
  {Khmel'Nitzkii}}, \bibinfo {author} {\bibfnamefont {A.}~\bibnamefont
  {Larkin}}, \ and\ \bibinfo {author} {\bibfnamefont {P.}~\bibnamefont {Lee}},\
  }\href@noop {} {\bibfield  {journal} {\bibinfo  {journal} {Physical Review
  B}\ }\textbf {\bibinfo {volume} {22}},\ \bibinfo {pages} {5142} (\bibinfo
  {year} {1980})}\BibitemShut {NoStop}%
\bibitem [{\citenamefont {Bergmann}(1984)}]{bergmann1984weak}%
  \BibitemOpen
  \bibfield  {author} {\bibinfo {author} {\bibfnamefont {G.}~\bibnamefont
  {Bergmann}},\ }\href@noop {} {\bibfield  {journal} {\bibinfo  {journal}
  {Physics Reports}\ }\textbf {\bibinfo {volume} {107}},\ \bibinfo {pages} {1}
  (\bibinfo {year} {1984})}\BibitemShut {NoStop}%
\bibitem [{\citenamefont {Licini}\ \emph {et~al.}(1985)\citenamefont {Licini},
  \citenamefont {Dolan},\ and\ \citenamefont {Bishop}}]{licini1985weakly}%
  \BibitemOpen
  \bibfield  {author} {\bibinfo {author} {\bibfnamefont {J.}~\bibnamefont
  {Licini}}, \bibinfo {author} {\bibfnamefont {G.}~\bibnamefont {Dolan}}, \
  and\ \bibinfo {author} {\bibfnamefont {D.}~\bibnamefont {Bishop}},\
  }\href@noop {} {\bibfield  {journal} {\bibinfo  {journal} {Physical review
  letters}\ }\textbf {\bibinfo {volume} {54}},\ \bibinfo {pages} {1585}
  (\bibinfo {year} {1985})}\BibitemShut {NoStop}%
\bibitem [{\citenamefont {Savchenko}\ \emph {et~al.}(1983)\citenamefont
  {Savchenko}, \citenamefont {Rylik},\ and\ \citenamefont
  {Lutskii}}]{savchenko1983antilocalization}%
  \BibitemOpen
  \bibfield  {author} {\bibinfo {author} {\bibfnamefont {A.}~\bibnamefont
  {Savchenko}}, \bibinfo {author} {\bibfnamefont {A.}~\bibnamefont {Rylik}}, \
  and\ \bibinfo {author} {\bibfnamefont {V.}~\bibnamefont {Lutskii}},\
  }\href@noop {} {\bibfield  {journal} {\bibinfo  {journal} {Zh. Eksp. Teor.
  Fiz.}\ }\textbf {\bibinfo {volume} {85}},\ \bibinfo {pages} {2210} (\bibinfo
  {year} {1983})}\BibitemShut {NoStop}%
\bibitem [{\citenamefont {Schierholz}\ \emph {et~al.}(2002)\citenamefont
  {Schierholz}, \citenamefont {K{\"u}rsten}, \citenamefont {Meier},
  \citenamefont {Matsuyama},\ and\ \citenamefont {Merkt}}]{schierholz2002weak}%
  \BibitemOpen
  \bibfield  {author} {\bibinfo {author} {\bibfnamefont {C.}~\bibnamefont
  {Schierholz}}, \bibinfo {author} {\bibfnamefont {R.}~\bibnamefont
  {K{\"u}rsten}}, \bibinfo {author} {\bibfnamefont {G.}~\bibnamefont {Meier}},
  \bibinfo {author} {\bibfnamefont {T.}~\bibnamefont {Matsuyama}}, \ and\
  \bibinfo {author} {\bibfnamefont {U.}~\bibnamefont {Merkt}},\ }\href@noop {}
  {\bibfield  {journal} {\bibinfo  {journal} {physica status solidi (b)}\
  }\textbf {\bibinfo {volume} {233}},\ \bibinfo {pages} {436} (\bibinfo {year}
  {2002})}\BibitemShut {NoStop}%
\bibitem [{\citenamefont {Miller}\ \emph {et~al.}(2003)\citenamefont {Miller},
  \citenamefont {Zumb{\"u}hl}, \citenamefont {Marcus}, \citenamefont
  {Lyanda-Geller}, \citenamefont {Goldhaber-Gordon}, \citenamefont {Campman},\
  and\ \citenamefont {Gossard}}]{miller2003gate}%
  \BibitemOpen
  \bibfield  {author} {\bibinfo {author} {\bibfnamefont {J.}~\bibnamefont
  {Miller}}, \bibinfo {author} {\bibfnamefont {D.}~\bibnamefont {Zumb{\"u}hl}},
  \bibinfo {author} {\bibfnamefont {C.}~\bibnamefont {Marcus}}, \bibinfo
  {author} {\bibfnamefont {Y.~B.}\ \bibnamefont {Lyanda-Geller}}, \bibinfo
  {author} {\bibfnamefont {D.}~\bibnamefont {Goldhaber-Gordon}}, \bibinfo
  {author} {\bibfnamefont {K.}~\bibnamefont {Campman}}, \ and\ \bibinfo
  {author} {\bibfnamefont {A.}~\bibnamefont {Gossard}},\ }\href@noop {}
  {\bibfield  {journal} {\bibinfo  {journal} {Physical review letters}\
  }\textbf {\bibinfo {volume} {90}},\ \bibinfo {pages} {076807} (\bibinfo
  {year} {2003})}\BibitemShut {NoStop}%
\bibitem [{\citenamefont {Nihei}\ \emph {et~al.}(2006)\citenamefont {Nihei},
  \citenamefont {Suzuki}, \citenamefont {Kohda},\ and\ \citenamefont
  {Nitta}}]{nihei2006gate}%
  \BibitemOpen
  \bibfield  {author} {\bibinfo {author} {\bibfnamefont {T.}~\bibnamefont
  {Nihei}}, \bibinfo {author} {\bibfnamefont {Y.}~\bibnamefont {Suzuki}},
  \bibinfo {author} {\bibfnamefont {M.}~\bibnamefont {Kohda}}, \ and\ \bibinfo
  {author} {\bibfnamefont {J.}~\bibnamefont {Nitta}},\ }\href@noop {}
  {\bibfield  {journal} {\bibinfo  {journal} {physica status solidi (c)}\
  }\textbf {\bibinfo {volume} {3}},\ \bibinfo {pages} {4239} (\bibinfo {year}
  {2006})}\BibitemShut {NoStop}%
\bibitem [{\citenamefont {Smorchkova}\ \emph {et~al.}(1997)\citenamefont
  {Smorchkova}, \citenamefont {Samarth}, \citenamefont {Kikkawa},\ and\
  \citenamefont {Awschalom}}]{smorchkova1997spin}%
  \BibitemOpen
  \bibfield  {author} {\bibinfo {author} {\bibfnamefont {I.}~\bibnamefont
  {Smorchkova}}, \bibinfo {author} {\bibfnamefont {N.}~\bibnamefont {Samarth}},
  \bibinfo {author} {\bibfnamefont {J.}~\bibnamefont {Kikkawa}}, \ and\
  \bibinfo {author} {\bibfnamefont {D.}~\bibnamefont {Awschalom}},\ }\href@noop
  {} {\bibfield  {journal} {\bibinfo  {journal} {Physical review letters}\
  }\textbf {\bibinfo {volume} {78}},\ \bibinfo {pages} {3571} (\bibinfo {year}
  {1997})}\BibitemShut {NoStop}%
\bibitem [{\citenamefont {Grbi{\'c}}\ \emph {et~al.}(2008)\citenamefont
  {Grbi{\'c}}, \citenamefont {Leturcq}, \citenamefont {Ihn}, \citenamefont
  {Ensslin}, \citenamefont {Reuter},\ and\ \citenamefont
  {Wieck}}]{grbic2008strong}%
  \BibitemOpen
  \bibfield  {author} {\bibinfo {author} {\bibfnamefont {B.}~\bibnamefont
  {Grbi{\'c}}}, \bibinfo {author} {\bibfnamefont {R.}~\bibnamefont {Leturcq}},
  \bibinfo {author} {\bibfnamefont {T.}~\bibnamefont {Ihn}}, \bibinfo {author}
  {\bibfnamefont {K.}~\bibnamefont {Ensslin}}, \bibinfo {author} {\bibfnamefont
  {D.}~\bibnamefont {Reuter}}, \ and\ \bibinfo {author} {\bibfnamefont {A.~D.}\
  \bibnamefont {Wieck}},\ }\href@noop {} {\bibfield  {journal} {\bibinfo
  {journal} {Physical Review B}\ }\textbf {\bibinfo {volume} {77}},\ \bibinfo
  {pages} {125312} (\bibinfo {year} {2008})}\BibitemShut {NoStop}%
\bibitem [{\citenamefont {Ando}\ \emph {et~al.}(1998)\citenamefont {Ando},
  \citenamefont {Nakanishi},\ and\ \citenamefont {Saito}}]{ando1998berry}%
  \BibitemOpen
  \bibfield  {author} {\bibinfo {author} {\bibfnamefont {T.}~\bibnamefont
  {Ando}}, \bibinfo {author} {\bibfnamefont {T.}~\bibnamefont {Nakanishi}}, \
  and\ \bibinfo {author} {\bibfnamefont {R.}~\bibnamefont {Saito}},\
  }\href@noop {} {\bibfield  {journal} {\bibinfo  {journal} {Journal of the
  Physical Society of Japan}\ }\textbf {\bibinfo {volume} {67}},\ \bibinfo
  {pages} {2857} (\bibinfo {year} {1998})}\BibitemShut {NoStop}%
\bibitem [{\citenamefont {Suzuura}\ and\ \citenamefont
  {Ando}(2002)}]{suzuura2002crossover}%
  \BibitemOpen
  \bibfield  {author} {\bibinfo {author} {\bibfnamefont {H.}~\bibnamefont
  {Suzuura}}\ and\ \bibinfo {author} {\bibfnamefont {T.}~\bibnamefont {Ando}},\
  }\href@noop {} {\bibfield  {journal} {\bibinfo  {journal} {Physical review
  letters}\ }\textbf {\bibinfo {volume} {89}},\ \bibinfo {pages} {266603}
  (\bibinfo {year} {2002})}\BibitemShut {NoStop}%
\bibitem [{\citenamefont {McCann}\ \emph {et~al.}(2006)\citenamefont {McCann},
  \citenamefont {Kechedzhi}, \citenamefont {Fal’ko}, \citenamefont {Suzuura},
  \citenamefont {Ando},\ and\ \citenamefont {Altshuler}}]{mccann2006weak}%
  \BibitemOpen
  \bibfield  {author} {\bibinfo {author} {\bibfnamefont {E.}~\bibnamefont
  {McCann}}, \bibinfo {author} {\bibfnamefont {K.}~\bibnamefont {Kechedzhi}},
  \bibinfo {author} {\bibfnamefont {V.~I.}\ \bibnamefont {Fal’ko}}, \bibinfo
  {author} {\bibfnamefont {H.}~\bibnamefont {Suzuura}}, \bibinfo {author}
  {\bibfnamefont {T.}~\bibnamefont {Ando}}, \ and\ \bibinfo {author}
  {\bibfnamefont {B.}~\bibnamefont {Altshuler}},\ }\href@noop {} {\bibfield
  {journal} {\bibinfo  {journal} {Physical Review Letters}\ }\textbf {\bibinfo
  {volume} {97}},\ \bibinfo {pages} {146805} (\bibinfo {year}
  {2006})}\BibitemShut {NoStop}%
\bibitem [{\citenamefont {Nomura}\ and\ \citenamefont
  {MacDonald}(2007)}]{nomura2007quantum}%
  \BibitemOpen
  \bibfield  {author} {\bibinfo {author} {\bibfnamefont {K.}~\bibnamefont
  {Nomura}}\ and\ \bibinfo {author} {\bibfnamefont {A.}~\bibnamefont
  {MacDonald}},\ }\href@noop {} {\bibfield  {journal} {\bibinfo  {journal}
  {Physical review letters}\ }\textbf {\bibinfo {volume} {98}},\ \bibinfo
  {pages} {076602} (\bibinfo {year} {2007})}\BibitemShut {NoStop}%
\bibitem [{\citenamefont {Morpurgo}\ and\ \citenamefont
  {Guinea}(2006)}]{morpurgo2006intervalley}%
  \BibitemOpen
  \bibfield  {author} {\bibinfo {author} {\bibfnamefont {A.}~\bibnamefont
  {Morpurgo}}\ and\ \bibinfo {author} {\bibfnamefont {F.}~\bibnamefont
  {Guinea}},\ }\href@noop {} {\bibfield  {journal} {\bibinfo  {journal}
  {Physical review letters}\ }\textbf {\bibinfo {volume} {97}},\ \bibinfo
  {pages} {196804} (\bibinfo {year} {2006})}\BibitemShut {NoStop}%
\bibitem [{\citenamefont {Kechedzhi}\ \emph {et~al.}(2007)\citenamefont
  {Kechedzhi}, \citenamefont {Fal’ko},\ and\ \citenamefont
  {McCann}}]{kechedzhi2007influence}%
  \BibitemOpen
  \bibfield  {author} {\bibinfo {author} {\bibfnamefont {K.}~\bibnamefont
  {Kechedzhi}}, \bibinfo {author} {\bibfnamefont {V.~I.}\ \bibnamefont
  {Fal’ko}}, \ and\ \bibinfo {author} {\bibfnamefont {B.}~\bibnamefont
  {McCann}, \bibfnamefont {E~anduler}},\ }\href@noop {} {\bibfield  {journal}
  {\bibinfo  {journal} {Physical review letters}\ }\textbf {\bibinfo {volume}
  {98}},\ \bibinfo {pages} {176806} (\bibinfo {year} {2007})}\BibitemShut
  {NoStop}%
\bibitem [{\citenamefont {Khveshchenko}(2006)}]{khveshchenko2006electron}%
  \BibitemOpen
  \bibfield  {author} {\bibinfo {author} {\bibfnamefont {D.}~\bibnamefont
  {Khveshchenko}},\ }\href@noop {} {\bibfield  {journal} {\bibinfo  {journal}
  {Physical review letters}\ }\textbf {\bibinfo {volume} {97}},\ \bibinfo
  {pages} {036802} (\bibinfo {year} {2006})}\BibitemShut {NoStop}%
\bibitem [{\citenamefont {Morozov}\ \emph {et~al.}(2006)\citenamefont
  {Morozov}, \citenamefont {Novoselov}, \citenamefont {Katsnelson},
  \citenamefont {Schedin}, \citenamefont {Ponomarenko}, \citenamefont {Jiang},\
  and\ \citenamefont {Geim}}]{morozov2006strong}%
  \BibitemOpen
  \bibfield  {author} {\bibinfo {author} {\bibfnamefont {S.}~\bibnamefont
  {Morozov}}, \bibinfo {author} {\bibfnamefont {K.}~\bibnamefont {Novoselov}},
  \bibinfo {author} {\bibfnamefont {M.}~\bibnamefont {Katsnelson}}, \bibinfo
  {author} {\bibfnamefont {F.}~\bibnamefont {Schedin}}, \bibinfo {author}
  {\bibfnamefont {L.}~\bibnamefont {Ponomarenko}}, \bibinfo {author}
  {\bibfnamefont {D.}~\bibnamefont {Jiang}}, \ and\ \bibinfo {author}
  {\bibfnamefont {A.}~\bibnamefont {Geim}},\ }\href@noop {} {\bibfield
  {journal} {\bibinfo  {journal} {Physical review letters}\ }\textbf {\bibinfo
  {volume} {97}},\ \bibinfo {pages} {016801} (\bibinfo {year}
  {2006})}\BibitemShut {NoStop}%
\bibitem [{\citenamefont {Heersche}\ \emph {et~al.}(2006)\citenamefont
  {Heersche}, \citenamefont {Jarillo-Herrero}, \citenamefont {Oostinga},
  \citenamefont {Vandersypen},\ and\ \citenamefont
  {Morpurgo}}]{heersche2006bipolar}%
  \BibitemOpen
  \bibfield  {author} {\bibinfo {author} {\bibfnamefont {H.~B.}\ \bibnamefont
  {Heersche}}, \bibinfo {author} {\bibfnamefont {P.}~\bibnamefont
  {Jarillo-Herrero}}, \bibinfo {author} {\bibfnamefont {J.~B.}\ \bibnamefont
  {Oostinga}}, \bibinfo {author} {\bibfnamefont {L.~M.}\ \bibnamefont
  {Vandersypen}}, \ and\ \bibinfo {author} {\bibfnamefont {A.~F.}\ \bibnamefont
  {Morpurgo}},\ }\href@noop {} {\bibfield  {journal} {\bibinfo  {journal}
  {arXiv preprint cond-mat/0612121}\ } (\bibinfo {year} {2006})}\BibitemShut
  {NoStop}%
\bibitem [{\citenamefont {Berger}\ \emph {et~al.}(2006)\citenamefont {Berger},
  \citenamefont {Song}, \citenamefont {Li}, \citenamefont {Wu}, \citenamefont
  {Brown}, \citenamefont {Naud}, \citenamefont {Mayou}, \citenamefont {Li},\
  and\ \citenamefont {Hass}}]{berger2006electronic}%
  \BibitemOpen
  \bibfield  {author} {\bibinfo {author} {\bibfnamefont {C.}~\bibnamefont
  {Berger}}, \bibinfo {author} {\bibfnamefont {Z.}~\bibnamefont {Song}},
  \bibinfo {author} {\bibfnamefont {X.}~\bibnamefont {Li}}, \bibinfo {author}
  {\bibfnamefont {X.}~\bibnamefont {Wu}}, \bibinfo {author} {\bibfnamefont
  {N.}~\bibnamefont {Brown}}, \bibinfo {author} {\bibfnamefont
  {C.}~\bibnamefont {Naud}}, \bibinfo {author} {\bibfnamefont {D.}~\bibnamefont
  {Mayou}}, \bibinfo {author} {\bibfnamefont {T.}~\bibnamefont {Li}}, \ and\
  \bibinfo {author} {\bibnamefont {Hass}},\ }\href@noop {} {\bibfield
  {journal} {\bibinfo  {journal} {Science}\ }\textbf {\bibinfo {volume}
  {312}},\ \bibinfo {pages} {1191} (\bibinfo {year} {2006})}\BibitemShut
  {NoStop}%
\bibitem [{\citenamefont {Tikhonenko}\ \emph {et~al.}(2009)\citenamefont
  {Tikhonenko}, \citenamefont {Kozikov}, \citenamefont {Savchenko},\ and\
  \citenamefont {Gorbachev}}]{tikhonenko2009transition}%
  \BibitemOpen
  \bibfield  {author} {\bibinfo {author} {\bibfnamefont {F.}~\bibnamefont
  {Tikhonenko}}, \bibinfo {author} {\bibfnamefont {A.}~\bibnamefont {Kozikov}},
  \bibinfo {author} {\bibfnamefont {A.}~\bibnamefont {Savchenko}}, \ and\
  \bibinfo {author} {\bibfnamefont {R.}~\bibnamefont {Gorbachev}},\ }\href@noop
  {} {\bibfield  {journal} {\bibinfo  {journal} {Physical Review Letters}\
  }\textbf {\bibinfo {volume} {103}},\ \bibinfo {pages} {226801} (\bibinfo
  {year} {2009})}\BibitemShut {NoStop}%
\bibitem [{\citenamefont {Berezovsky}\ and\ \citenamefont
  {Westervelt}(2010)}]{berezovsky2010imaging}%
  \BibitemOpen
  \bibfield  {author} {\bibinfo {author} {\bibfnamefont {J.}~\bibnamefont
  {Berezovsky}}\ and\ \bibinfo {author} {\bibfnamefont {R.~M.}\ \bibnamefont
  {Westervelt}},\ }\href@noop {} {\  (\bibinfo {year} {2010})}\BibitemShut
  {NoStop}%
\bibitem [{\citenamefont {Chen}\ \emph {et~al.}(2010)\citenamefont {Chen},
  \citenamefont {Bae}, \citenamefont {Chialvo}, \citenamefont {Dirks},
  \citenamefont {Bezryadin},\ and\ \citenamefont
  {Mason}}]{chen2010magnetoresistance}%
  \BibitemOpen
  \bibfield  {author} {\bibinfo {author} {\bibfnamefont {Y.-F.}\ \bibnamefont
  {Chen}}, \bibinfo {author} {\bibfnamefont {M.-H.}\ \bibnamefont {Bae}},
  \bibinfo {author} {\bibfnamefont {C.}~\bibnamefont {Chialvo}}, \bibinfo
  {author} {\bibfnamefont {T.}~\bibnamefont {Dirks}}, \bibinfo {author}
  {\bibfnamefont {A.}~\bibnamefont {Bezryadin}}, \ and\ \bibinfo {author}
  {\bibfnamefont {N.}~\bibnamefont {Mason}},\ }\href@noop {} {\bibfield
  {journal} {\bibinfo  {journal} {Journal of Physics: Condensed Matter}\
  }\textbf {\bibinfo {volume} {22}},\ \bibinfo {pages} {205301} (\bibinfo
  {year} {2010})}\BibitemShut {NoStop}%
\bibitem [{\citenamefont {Baker}\ \emph {et~al.}(2012)\citenamefont {Baker},
  \citenamefont {Alexander-Webber}, \citenamefont {Altebaeumer}, \citenamefont
  {Janssen}, \citenamefont {Tzalenchuk}, \citenamefont {Lara-Avila},
  \citenamefont {Kubatkin},\ and\ \citenamefont {Yakimova}}]{baker2012weak}%
  \BibitemOpen
  \bibfield  {author} {\bibinfo {author} {\bibfnamefont {A.}~\bibnamefont
  {Baker}}, \bibinfo {author} {\bibfnamefont {J.}~\bibnamefont
  {Alexander-Webber}}, \bibinfo {author} {\bibfnamefont {T.}~\bibnamefont
  {Altebaeumer}}, \bibinfo {author} {\bibfnamefont {T.}~\bibnamefont
  {Janssen}}, \bibinfo {author} {\bibfnamefont {A.}~\bibnamefont {Tzalenchuk}},
  \bibinfo {author} {\bibfnamefont {S.}~\bibnamefont {Lara-Avila}}, \bibinfo
  {author} {\bibfnamefont {S.}~\bibnamefont {Kubatkin}}, \ and\ \bibinfo
  {author} {\bibnamefont {Yakimova}},\ }\href@noop {} {\bibfield  {journal}
  {\bibinfo  {journal} {Physical Review B}\ }\textbf {\bibinfo {volume} {86}},\
  \bibinfo {pages} {235441} (\bibinfo {year} {2012})}\BibitemShut {NoStop}%
\bibitem [{\citenamefont {Mahmood}\ \emph {et~al.}(2013)\citenamefont
  {Mahmood}, \citenamefont {Naud}, \citenamefont {Bouvier}, \citenamefont
  {Hiebel}, \citenamefont {Mallet}, \citenamefont {Veuillen}, \citenamefont
  {L{\'e}vy}, \citenamefont {Chaussende},\ and\ \citenamefont
  {Ouisse}}]{mahmood2013epitaxial}%
  \BibitemOpen
  \bibfield  {author} {\bibinfo {author} {\bibfnamefont {A.}~\bibnamefont
  {Mahmood}}, \bibinfo {author} {\bibfnamefont {C.}~\bibnamefont {Naud}},
  \bibinfo {author} {\bibfnamefont {C.}~\bibnamefont {Bouvier}}, \bibinfo
  {author} {\bibfnamefont {F.}~\bibnamefont {Hiebel}}, \bibinfo {author}
  {\bibfnamefont {P.}~\bibnamefont {Mallet}}, \bibinfo {author} {\bibfnamefont
  {J.-Y.}\ \bibnamefont {Veuillen}}, \bibinfo {author} {\bibfnamefont
  {L.}~\bibnamefont {L{\'e}vy}}, \bibinfo {author} {\bibfnamefont
  {D.}~\bibnamefont {Chaussende}}, \ and\ \bibinfo {author} {\bibfnamefont
  {T.}~\bibnamefont {Ouisse}},\ }\href@noop {} {\bibfield  {journal} {\bibinfo
  {journal} {Journal of Applied Physics}\ }\textbf {\bibinfo {volume} {113}},\
  \bibinfo {pages} {083715} (\bibinfo {year} {2013})}\BibitemShut {NoStop}%
\bibitem [{\citenamefont {Iqbal}\ \emph {et~al.}(2014)\citenamefont {Iqbal},
  \citenamefont {Kelek{\c{c}}i}, \citenamefont {Iqbal}, \citenamefont {Jin},
  \citenamefont {Hwang},\ and\ \citenamefont {Eom}}]{iqbal2014enhanced}%
  \BibitemOpen
  \bibfield  {author} {\bibinfo {author} {\bibfnamefont {M.}~\bibnamefont
  {Iqbal}}, \bibinfo {author} {\bibfnamefont {{\"O}.}~\bibnamefont
  {Kelek{\c{c}}i}}, \bibinfo {author} {\bibfnamefont {M.}~\bibnamefont
  {Iqbal}}, \bibinfo {author} {\bibfnamefont {X.}~\bibnamefont {Jin}}, \bibinfo
  {author} {\bibfnamefont {C.}~\bibnamefont {Hwang}}, \ and\ \bibinfo {author}
  {\bibfnamefont {J.}~\bibnamefont {Eom}},\ }\href@noop {} {\bibfield
  {journal} {\bibinfo  {journal} {New Journal of Physics}\ }\textbf {\bibinfo
  {volume} {16}},\ \bibinfo {pages} {083020} (\bibinfo {year}
  {2014})}\BibitemShut {NoStop}%
\bibitem [{\citenamefont {Hilke}\ \emph {et~al.}(2014)\citenamefont {Hilke},
  \citenamefont {Massicotte}, \citenamefont {Whiteway},\ and\ \citenamefont
  {Yu}}]{hilke2014weak}%
  \BibitemOpen
  \bibfield  {author} {\bibinfo {author} {\bibfnamefont {M.}~\bibnamefont
  {Hilke}}, \bibinfo {author} {\bibfnamefont {M.}~\bibnamefont {Massicotte}},
  \bibinfo {author} {\bibfnamefont {E.}~\bibnamefont {Whiteway}}, \ and\
  \bibinfo {author} {\bibfnamefont {V.}~\bibnamefont {Yu}},\ }\href@noop {}
  {\bibfield  {journal} {\bibinfo  {journal} {The Scientific World Journal}\
  }\textbf {\bibinfo {volume} {2014}} (\bibinfo {year} {2014})}\BibitemShut
  {NoStop}%
\bibitem [{\citenamefont {Chandni}\ \emph {et~al.}(2015)\citenamefont
  {Chandni}, \citenamefont {Henriksen},\ and\ \citenamefont
  {Eisenstein}}]{chandni2015transport}%
  \BibitemOpen
  \bibfield  {author} {\bibinfo {author} {\bibfnamefont {U.}~\bibnamefont
  {Chandni}}, \bibinfo {author} {\bibfnamefont {E.~A.}\ \bibnamefont
  {Henriksen}}, \ and\ \bibinfo {author} {\bibfnamefont {J.}~\bibnamefont
  {Eisenstein}},\ }\href@noop {} {\bibfield  {journal} {\bibinfo  {journal}
  {Physical Review B}\ }\textbf {\bibinfo {volume} {91}},\ \bibinfo {pages}
  {245402} (\bibinfo {year} {2015})}\BibitemShut {NoStop}%
\bibitem [{\citenamefont {Pal}\ \emph {et~al.}(2012)\citenamefont {Pal},
  \citenamefont {Kochat},\ and\ \citenamefont {Ghosh}}]{pal2012direct}%
  \BibitemOpen
  \bibfield  {author} {\bibinfo {author} {\bibfnamefont {A.~N.}\ \bibnamefont
  {Pal}}, \bibinfo {author} {\bibfnamefont {V.}~\bibnamefont {Kochat}}, \ and\
  \bibinfo {author} {\bibfnamefont {A.}~\bibnamefont {Ghosh}},\ }\href@noop {}
  {\bibfield  {journal} {\bibinfo  {journal} {Physical Review Letters}\
  }\textbf {\bibinfo {volume} {109}},\ \bibinfo {pages} {196601} (\bibinfo
  {year} {2012})}\BibitemShut {NoStop}%
\bibitem [{\citenamefont {Ki}\ \emph {et~al.}(2008)\citenamefont {Ki},
  \citenamefont {Jeong}, \citenamefont {Choi}, \citenamefont {Lee},\ and\
  \citenamefont {Park}}]{ki2008inelastic}%
  \BibitemOpen
  \bibfield  {author} {\bibinfo {author} {\bibfnamefont {D.-K.}\ \bibnamefont
  {Ki}}, \bibinfo {author} {\bibfnamefont {D.}~\bibnamefont {Jeong}}, \bibinfo
  {author} {\bibfnamefont {J.-H.}\ \bibnamefont {Choi}}, \bibinfo {author}
  {\bibfnamefont {H.-J.}\ \bibnamefont {Lee}}, \ and\ \bibinfo {author}
  {\bibfnamefont {K.-S.}\ \bibnamefont {Park}},\ }\href@noop {} {\bibfield
  {journal} {\bibinfo  {journal} {Physical Review B}\ }\textbf {\bibinfo
  {volume} {78}},\ \bibinfo {pages} {125409} (\bibinfo {year}
  {2008})}\BibitemShut {NoStop}%
\bibitem [{\citenamefont {Tikhonenko}\ \emph {et~al.}(2008)\citenamefont
  {Tikhonenko}, \citenamefont {Horsell}, \citenamefont {Gorbachev},\ and\
  \citenamefont {Savchenko}}]{tikhonenko2008weak}%
  \BibitemOpen
  \bibfield  {author} {\bibinfo {author} {\bibfnamefont {F.}~\bibnamefont
  {Tikhonenko}}, \bibinfo {author} {\bibfnamefont {D.}~\bibnamefont {Horsell}},
  \bibinfo {author} {\bibfnamefont {R.}~\bibnamefont {Gorbachev}}, \ and\
  \bibinfo {author} {\bibfnamefont {A.}~\bibnamefont {Savchenko}},\ }\href@noop
  {} {\bibfield  {journal} {\bibinfo  {journal} {Physical review letters}\
  }\textbf {\bibinfo {volume} {100}},\ \bibinfo {pages} {056802} (\bibinfo
  {year} {2008})}\BibitemShut {NoStop}%
\bibitem [{\citenamefont {Oh}\ \emph {et~al.}(2010)\citenamefont {Oh},
  \citenamefont {Eom}, \citenamefont {Koo},\ and\ \citenamefont
  {Han}}]{oh2010electronic}%
  \BibitemOpen
  \bibfield  {author} {\bibinfo {author} {\bibfnamefont {Y.}~\bibnamefont
  {Oh}}, \bibinfo {author} {\bibfnamefont {J.}~\bibnamefont {Eom}}, \bibinfo
  {author} {\bibfnamefont {H.~C.}\ \bibnamefont {Koo}}, \ and\ \bibinfo
  {author} {\bibfnamefont {S.~H.}\ \bibnamefont {Han}},\ }\href@noop {}
  {\bibfield  {journal} {\bibinfo  {journal} {Solid State Communications}\
  }\textbf {\bibinfo {volume} {150}},\ \bibinfo {pages} {1987} (\bibinfo {year}
  {2010})}\BibitemShut {NoStop}%
\bibitem [{\citenamefont {Moser}\ \emph {et~al.}(2010)\citenamefont {Moser},
  \citenamefont {Tao}, \citenamefont {Roche}, \citenamefont {Alzina},
  \citenamefont {Torres},\ and\ \citenamefont
  {Bachtold}}]{moser2010magnetotransport}%
  \BibitemOpen
  \bibfield  {author} {\bibinfo {author} {\bibfnamefont {J.}~\bibnamefont
  {Moser}}, \bibinfo {author} {\bibfnamefont {H.}~\bibnamefont {Tao}}, \bibinfo
  {author} {\bibfnamefont {S.}~\bibnamefont {Roche}}, \bibinfo {author}
  {\bibfnamefont {F.}~\bibnamefont {Alzina}}, \bibinfo {author} {\bibfnamefont
  {C.~S.}\ \bibnamefont {Torres}}, \ and\ \bibinfo {author} {\bibfnamefont
  {A.}~\bibnamefont {Bachtold}},\ }\href@noop {} {\bibfield  {journal}
  {\bibinfo  {journal} {Physical Review B}\ }\textbf {\bibinfo {volume} {81}},\
  \bibinfo {pages} {205445} (\bibinfo {year} {2010})}\BibitemShut {NoStop}%
\bibitem [{\citenamefont {Lundeberg}\ and\ \citenamefont
  {Folk}(2010)}]{lundeberg2010rippled}%
  \BibitemOpen
  \bibfield  {author} {\bibinfo {author} {\bibfnamefont {M.~B.}\ \bibnamefont
  {Lundeberg}}\ and\ \bibinfo {author} {\bibfnamefont {J.~A.}\ \bibnamefont
  {Folk}},\ }\href@noop {} {\bibfield  {journal} {\bibinfo  {journal} {Physical
  review letters}\ }\textbf {\bibinfo {volume} {105}},\ \bibinfo {pages}
  {146804} (\bibinfo {year} {2010})}\BibitemShut {NoStop}%
\bibitem [{\citenamefont {Wu}\ \emph {et~al.}(2007)\citenamefont {Wu},
  \citenamefont {Li}, \citenamefont {Song}, \citenamefont {Berger},\ and\
  \citenamefont {de~Heer}}]{wu2007weak}%
  \BibitemOpen
  \bibfield  {author} {\bibinfo {author} {\bibfnamefont {X.}~\bibnamefont
  {Wu}}, \bibinfo {author} {\bibfnamefont {X.}~\bibnamefont {Li}}, \bibinfo
  {author} {\bibfnamefont {Z.}~\bibnamefont {Song}}, \bibinfo {author}
  {\bibfnamefont {C.}~\bibnamefont {Berger}}, \ and\ \bibinfo {author}
  {\bibfnamefont {W.~A.}\ \bibnamefont {de~Heer}},\ }\href@noop {} {\bibfield
  {journal} {\bibinfo  {journal} {Physical review letters}\ }\textbf {\bibinfo
  {volume} {98}},\ \bibinfo {pages} {136801} (\bibinfo {year}
  {2007})}\BibitemShut {NoStop}%
\bibitem [{\citenamefont {Liu}\ \emph {et~al.}(2011)\citenamefont {Liu},
  \citenamefont {Lew},\ and\ \citenamefont {Sun}}]{liu2011enhanced}%
  \BibitemOpen
  \bibfield  {author} {\bibinfo {author} {\bibfnamefont {Y.}~\bibnamefont
  {Liu}}, \bibinfo {author} {\bibfnamefont {W.~S.}\ \bibnamefont {Lew}}, \ and\
  \bibinfo {author} {\bibfnamefont {L.}~\bibnamefont {Sun}},\ }\href@noop {}
  {\bibfield  {journal} {\bibinfo  {journal} {Physical Chemistry Chemical
  Physics}\ }\textbf {\bibinfo {volume} {13}},\ \bibinfo {pages} {20208}
  (\bibinfo {year} {2011})}\BibitemShut {NoStop}%
\bibitem [{\citenamefont {Jouault}\ \emph {et~al.}(2011)\citenamefont
  {Jouault}, \citenamefont {Jabakhanji}, \citenamefont {Camara}, \citenamefont
  {Desrat}, \citenamefont {Consejo},\ and\ \citenamefont
  {Camassel}}]{jouault2011interplay}%
  \BibitemOpen
  \bibfield  {author} {\bibinfo {author} {\bibfnamefont {B.}~\bibnamefont
  {Jouault}}, \bibinfo {author} {\bibfnamefont {B.}~\bibnamefont {Jabakhanji}},
  \bibinfo {author} {\bibfnamefont {N.}~\bibnamefont {Camara}}, \bibinfo
  {author} {\bibfnamefont {W.}~\bibnamefont {Desrat}}, \bibinfo {author}
  {\bibfnamefont {C.}~\bibnamefont {Consejo}}, \ and\ \bibinfo {author}
  {\bibfnamefont {J.}~\bibnamefont {Camassel}},\ }\href@noop {} {\bibfield
  {journal} {\bibinfo  {journal} {Physical Review B}\ }\textbf {\bibinfo
  {volume} {83}},\ \bibinfo {pages} {195417} (\bibinfo {year}
  {2011})}\BibitemShut {NoStop}%
\bibitem [{\citenamefont {Fal’ko}\ \emph {et~al.}(2007)\citenamefont
  {Fal’ko}, \citenamefont {Kechedzhi}, \citenamefont {McCann}, \citenamefont
  {Altshuler}, \citenamefont {Suzuura},\ and\ \citenamefont
  {Ando}}]{fal2007weak}%
  \BibitemOpen
  \bibfield  {author} {\bibinfo {author} {\bibfnamefont {V.~I.}\ \bibnamefont
  {Fal’ko}}, \bibinfo {author} {\bibfnamefont {K.}~\bibnamefont {Kechedzhi}},
  \bibinfo {author} {\bibfnamefont {E.}~\bibnamefont {McCann}}, \bibinfo
  {author} {\bibfnamefont {B.}~\bibnamefont {Altshuler}}, \bibinfo {author}
  {\bibfnamefont {H.}~\bibnamefont {Suzuura}}, \ and\ \bibinfo {author}
  {\bibfnamefont {T.}~\bibnamefont {Ando}},\ }\href@noop {} {\bibfield
  {journal} {\bibinfo  {journal} {Solid State Communications}\ }\textbf
  {\bibinfo {volume} {143}},\ \bibinfo {pages} {33} (\bibinfo {year}
  {2007})}\BibitemShut {NoStop}%
\bibitem [{\citenamefont {Chen}\ \emph {et~al.}(2011)\citenamefont {Chen},
  \citenamefont {Bae}, \citenamefont {Chialvo}, \citenamefont {Dirks},
  \citenamefont {Bezryadin},\ and\ \citenamefont {Mason}}]{chen2011negative}%
  \BibitemOpen
  \bibfield  {author} {\bibinfo {author} {\bibfnamefont {Y.-F.}\ \bibnamefont
  {Chen}}, \bibinfo {author} {\bibfnamefont {M.-H.}\ \bibnamefont {Bae}},
  \bibinfo {author} {\bibfnamefont {C.}~\bibnamefont {Chialvo}}, \bibinfo
  {author} {\bibfnamefont {T.}~\bibnamefont {Dirks}}, \bibinfo {author}
  {\bibfnamefont {A.}~\bibnamefont {Bezryadin}}, \ and\ \bibinfo {author}
  {\bibfnamefont {N.}~\bibnamefont {Mason}},\ }\href@noop {} {\bibfield
  {journal} {\bibinfo  {journal} {Physica B: Condensed Matter}\ }\textbf
  {\bibinfo {volume} {406}},\ \bibinfo {pages} {785} (\bibinfo {year}
  {2011})}\BibitemShut {NoStop}%
\bibitem [{\citenamefont {Aleiner}\ and\ \citenamefont
  {Efetov}(2006)}]{aleiner2006effect}%
  \BibitemOpen
  \bibfield  {author} {\bibinfo {author} {\bibfnamefont {I.}~\bibnamefont
  {Aleiner}}\ and\ \bibinfo {author} {\bibfnamefont {K.}~\bibnamefont
  {Efetov}},\ }\href@noop {} {\bibfield  {journal} {\bibinfo  {journal}
  {Physical review letters}\ }\textbf {\bibinfo {volume} {97}},\ \bibinfo
  {pages} {236801} (\bibinfo {year} {2006})}\BibitemShut {NoStop}%
\bibitem [{\citenamefont {Wang}\ \emph
  {et~al.}(2015{\natexlab{a}})\citenamefont {Wang}, \citenamefont {Cheng},
  \citenamefont {Martynov}, \citenamefont {Miao}, \citenamefont {Jing},
  \citenamefont {Taniguchi}, \citenamefont {Watanabe}, \citenamefont {Aji},
  \citenamefont {Lau},\ and\ \citenamefont {Bockrath}}]{wang2015topological}%
  \BibitemOpen
  \bibfield  {author} {\bibinfo {author} {\bibfnamefont {P.}~\bibnamefont
  {Wang}}, \bibinfo {author} {\bibfnamefont {B.}~\bibnamefont {Cheng}},
  \bibinfo {author} {\bibfnamefont {O.}~\bibnamefont {Martynov}}, \bibinfo
  {author} {\bibfnamefont {T.}~\bibnamefont {Miao}}, \bibinfo {author}
  {\bibfnamefont {L.}~\bibnamefont {Jing}}, \bibinfo {author} {\bibfnamefont
  {T.}~\bibnamefont {Taniguchi}}, \bibinfo {author} {\bibfnamefont
  {K.}~\bibnamefont {Watanabe}}, \bibinfo {author} {\bibfnamefont
  {V.}~\bibnamefont {Aji}}, \bibinfo {author} {\bibfnamefont {C.~N.}\
  \bibnamefont {Lau}}, \ and\ \bibinfo {author} {\bibfnamefont
  {M.}~\bibnamefont {Bockrath}},\ }\href@noop {} {\bibfield  {journal}
  {\bibinfo  {journal} {Nano Letters}\ }\textbf {\bibinfo {volume} {15}},\
  \bibinfo {pages} {6395} (\bibinfo {year} {2015}{\natexlab{a}})}\BibitemShut
  {NoStop}%
\bibitem [{\citenamefont {Zomer}\ \emph {et~al.}(2011)\citenamefont {Zomer},
  \citenamefont {Dash}, \citenamefont {Tombros},\ and\ \citenamefont
  {Van~Wees}}]{zomer2011transfer}%
  \BibitemOpen
  \bibfield  {author} {\bibinfo {author} {\bibfnamefont {P.}~\bibnamefont
  {Zomer}}, \bibinfo {author} {\bibfnamefont {S.}~\bibnamefont {Dash}},
  \bibinfo {author} {\bibfnamefont {N.}~\bibnamefont {Tombros}}, \ and\
  \bibinfo {author} {\bibfnamefont {B.}~\bibnamefont {Van~Wees}},\ }\href@noop
  {} {\bibfield  {journal} {\bibinfo  {journal} {Applied Physics Letters}\
  }\textbf {\bibinfo {volume} {99}},\ \bibinfo {pages} {232104} (\bibinfo
  {year} {2011})}\BibitemShut {NoStop}%
\bibitem [{\citenamefont {Yankowitz}\ \emph {et~al.}(2012)\citenamefont
  {Yankowitz}, \citenamefont {Xue}, \citenamefont {Cormode}, \citenamefont
  {Sanchez-Yamagishi}, \citenamefont {Watanabe}, \citenamefont {Taniguchi},
  \citenamefont {Jarillo-Herrero}, \citenamefont {Jacquod},\ and\ \citenamefont
  {LeRoy}}]{yankowitz2012emergence}%
  \BibitemOpen
  \bibfield  {author} {\bibinfo {author} {\bibfnamefont {M.}~\bibnamefont
  {Yankowitz}}, \bibinfo {author} {\bibfnamefont {J.}~\bibnamefont {Xue}},
  \bibinfo {author} {\bibfnamefont {D.}~\bibnamefont {Cormode}}, \bibinfo
  {author} {\bibfnamefont {J.~D.}\ \bibnamefont {Sanchez-Yamagishi}}, \bibinfo
  {author} {\bibfnamefont {K.}~\bibnamefont {Watanabe}}, \bibinfo {author}
  {\bibfnamefont {T.}~\bibnamefont {Taniguchi}}, \bibinfo {author}
  {\bibfnamefont {P.}~\bibnamefont {Jarillo-Herrero}}, \bibinfo {author}
  {\bibfnamefont {P.}~\bibnamefont {Jacquod}}, \ and\ \bibinfo {author}
  {\bibfnamefont {B.~J.}\ \bibnamefont {LeRoy}},\ }\href@noop {} {\bibfield
  {journal} {\bibinfo  {journal} {Nature Physics}\ }\textbf {\bibinfo {volume}
  {8}},\ \bibinfo {pages} {382} (\bibinfo {year} {2012})}\BibitemShut {NoStop}%
\bibitem [{\citenamefont {Hunt}\ \emph {et~al.}(2013)\citenamefont {Hunt},
  \citenamefont {Sanchez-Yamagishi}, \citenamefont {Young}, \citenamefont
  {Yankowitz}, \citenamefont {LeRoy}, \citenamefont {Watanabe}, \citenamefont
  {Taniguchi}, \citenamefont {Moon}, \citenamefont {Koshino},\ and\
  \citenamefont {Jarillo-Herrero}}]{hunt2013massive}%
  \BibitemOpen
  \bibfield  {author} {\bibinfo {author} {\bibfnamefont {B.}~\bibnamefont
  {Hunt}}, \bibinfo {author} {\bibfnamefont {J.}~\bibnamefont
  {Sanchez-Yamagishi}}, \bibinfo {author} {\bibfnamefont {A.}~\bibnamefont
  {Young}}, \bibinfo {author} {\bibfnamefont {M.}~\bibnamefont {Yankowitz}},
  \bibinfo {author} {\bibfnamefont {B.~J.}\ \bibnamefont {LeRoy}}, \bibinfo
  {author} {\bibfnamefont {K.}~\bibnamefont {Watanabe}}, \bibinfo {author}
  {\bibfnamefont {T.}~\bibnamefont {Taniguchi}}, \bibinfo {author}
  {\bibfnamefont {P.}~\bibnamefont {Moon}}, \bibinfo {author} {\bibfnamefont
  {M.}~\bibnamefont {Koshino}}, \ and\ \bibinfo {author} {\bibnamefont
  {Jarillo-Herrero}},\ }\href@noop {} {\bibfield  {journal} {\bibinfo
  {journal} {Science}\ }\textbf {\bibinfo {volume} {340}},\ \bibinfo {pages}
  {1427} (\bibinfo {year} {2013})}\BibitemShut {NoStop}%
\bibitem [{\citenamefont {Dean}\ \emph {et~al.}(2013)\citenamefont {Dean},
  \citenamefont {Wang}, \citenamefont {Maher}, \citenamefont {Forsythe},
  \citenamefont {Ghahari}, \citenamefont {Gao}, \citenamefont {Katoch},
  \citenamefont {Ishigami}, \citenamefont {Moon},\ and\ \citenamefont
  {Koshino}}]{dean2013hofstadter}%
  \BibitemOpen
  \bibfield  {author} {\bibinfo {author} {\bibfnamefont {C.}~\bibnamefont
  {Dean}}, \bibinfo {author} {\bibfnamefont {L.}~\bibnamefont {Wang}}, \bibinfo
  {author} {\bibfnamefont {P.}~\bibnamefont {Maher}}, \bibinfo {author}
  {\bibfnamefont {C.}~\bibnamefont {Forsythe}}, \bibinfo {author}
  {\bibfnamefont {F.}~\bibnamefont {Ghahari}}, \bibinfo {author} {\bibfnamefont
  {Y.}~\bibnamefont {Gao}}, \bibinfo {author} {\bibfnamefont {J.}~\bibnamefont
  {Katoch}}, \bibinfo {author} {\bibfnamefont {M.}~\bibnamefont {Ishigami}},
  \bibinfo {author} {\bibfnamefont {P.}~\bibnamefont {Moon}}, \ and\ \bibinfo
  {author} {\bibnamefont {Koshino}},\ }\href@noop {} {\bibfield  {journal}
  {\bibinfo  {journal} {Nature}\ }\textbf {\bibinfo {volume} {497}},\ \bibinfo
  {pages} {598} (\bibinfo {year} {2013})}\BibitemShut {NoStop}%
\bibitem [{\citenamefont {Ponomarenko}\ \emph {et~al.}(2013)\citenamefont
  {Ponomarenko}, \citenamefont {Gorbachev}, \citenamefont {Yu}, \citenamefont
  {Elias}, \citenamefont {Jalil}, \citenamefont {Patel}, \citenamefont
  {Mishchenko}, \citenamefont {Mayorov}, \citenamefont {Woods},\ and\
  \citenamefont {Wallbank}}]{ponomarenko2013cloning}%
  \BibitemOpen
  \bibfield  {author} {\bibinfo {author} {\bibfnamefont {L.}~\bibnamefont
  {Ponomarenko}}, \bibinfo {author} {\bibfnamefont {R.}~\bibnamefont
  {Gorbachev}}, \bibinfo {author} {\bibfnamefont {G.}~\bibnamefont {Yu}},
  \bibinfo {author} {\bibfnamefont {D.}~\bibnamefont {Elias}}, \bibinfo
  {author} {\bibfnamefont {R.}~\bibnamefont {Jalil}}, \bibinfo {author}
  {\bibfnamefont {A.}~\bibnamefont {Patel}}, \bibinfo {author} {\bibfnamefont
  {A.}~\bibnamefont {Mishchenko}}, \bibinfo {author} {\bibfnamefont
  {A.}~\bibnamefont {Mayorov}}, \bibinfo {author} {\bibfnamefont
  {C.}~\bibnamefont {Woods}}, \ and\ \bibinfo {author} {\bibnamefont
  {Wallbank}},\ }\href@noop {} {\bibfield  {journal} {\bibinfo  {journal}
  {Nature}\ }\textbf {\bibinfo {volume} {497}},\ \bibinfo {pages} {594}
  (\bibinfo {year} {2013})}\BibitemShut {NoStop}%
\bibitem [{\citenamefont {Yu}\ \emph {et~al.}(2014)\citenamefont {Yu},
  \citenamefont {Gorbachev}, \citenamefont {Tu}, \citenamefont {Kretinin},
  \citenamefont {Cao}, \citenamefont {Jalil}, \citenamefont {Withers},
  \citenamefont {Ponomarenko}, \citenamefont {Piot},\ and\ \citenamefont
  {Potemski}}]{yu2014hierarchy}%
  \BibitemOpen
  \bibfield  {author} {\bibinfo {author} {\bibfnamefont {G.}~\bibnamefont
  {Yu}}, \bibinfo {author} {\bibfnamefont {R.}~\bibnamefont {Gorbachev}},
  \bibinfo {author} {\bibfnamefont {J.}~\bibnamefont {Tu}}, \bibinfo {author}
  {\bibfnamefont {A.}~\bibnamefont {Kretinin}}, \bibinfo {author}
  {\bibfnamefont {Y.}~\bibnamefont {Cao}}, \bibinfo {author} {\bibfnamefont
  {R.}~\bibnamefont {Jalil}}, \bibinfo {author} {\bibfnamefont
  {F.}~\bibnamefont {Withers}}, \bibinfo {author} {\bibfnamefont
  {L.}~\bibnamefont {Ponomarenko}}, \bibinfo {author} {\bibfnamefont
  {B.}~\bibnamefont {Piot}}, \ and\ \bibinfo {author} {\bibnamefont
  {Potemski}},\ }\href@noop {} {\bibfield  {journal} {\bibinfo  {journal}
  {Nature physics}\ }\textbf {\bibinfo {volume} {10}},\ \bibinfo {pages} {525}
  (\bibinfo {year} {2014})}\BibitemShut {NoStop}%
\bibitem [{\citenamefont {Woods}\ \emph {et~al.}(2014)\citenamefont {Woods},
  \citenamefont {Britnell}, \citenamefont {Eckmann}, \citenamefont {Ma},
  \citenamefont {Lu}, \citenamefont {Guo}, \citenamefont {Lin}, \citenamefont
  {Yu}, \citenamefont {Cao},\ and\ \citenamefont
  {Gorbachev}}]{woods2014commensurate}%
  \BibitemOpen
  \bibfield  {author} {\bibinfo {author} {\bibfnamefont {C.}~\bibnamefont
  {Woods}}, \bibinfo {author} {\bibfnamefont {L.}~\bibnamefont {Britnell}},
  \bibinfo {author} {\bibfnamefont {A.}~\bibnamefont {Eckmann}}, \bibinfo
  {author} {\bibfnamefont {R.}~\bibnamefont {Ma}}, \bibinfo {author}
  {\bibfnamefont {J.}~\bibnamefont {Lu}}, \bibinfo {author} {\bibfnamefont
  {H.}~\bibnamefont {Guo}}, \bibinfo {author} {\bibfnamefont {X.}~\bibnamefont
  {Lin}}, \bibinfo {author} {\bibfnamefont {G.}~\bibnamefont {Yu}}, \bibinfo
  {author} {\bibfnamefont {Y.}~\bibnamefont {Cao}}, \ and\ \bibinfo {author}
  {\bibnamefont {Gorbachev}},\ }\href@noop {} {\bibfield  {journal} {\bibinfo
  {journal} {Nature physics}\ }\textbf {\bibinfo {volume} {10}},\ \bibinfo
  {pages} {451} (\bibinfo {year} {2014})}\BibitemShut {NoStop}%
\bibitem [{\citenamefont {Gorbachev}\ \emph {et~al.}(2014)\citenamefont
  {Gorbachev}, \citenamefont {Song}, \citenamefont {Yu}, \citenamefont
  {Kretinin}, \citenamefont {Withers}, \citenamefont {Cao}, \citenamefont
  {Mishchenko}, \citenamefont {Grigorieva},\ and\ \citenamefont
  {Novoselov}}]{gorbachev2014detecting}%
  \BibitemOpen
  \bibfield  {author} {\bibinfo {author} {\bibfnamefont {R.}~\bibnamefont
  {Gorbachev}}, \bibinfo {author} {\bibfnamefont {J.}~\bibnamefont {Song}},
  \bibinfo {author} {\bibfnamefont {G.}~\bibnamefont {Yu}}, \bibinfo {author}
  {\bibfnamefont {A.}~\bibnamefont {Kretinin}}, \bibinfo {author}
  {\bibfnamefont {F.}~\bibnamefont {Withers}}, \bibinfo {author} {\bibfnamefont
  {Y.}~\bibnamefont {Cao}}, \bibinfo {author} {\bibfnamefont {A.}~\bibnamefont
  {Mishchenko}}, \bibinfo {author} {\bibfnamefont {I.}~\bibnamefont
  {Grigorieva}}, \ and\ \bibinfo {author} {\bibnamefont {Novoselov}},\
  }\href@noop {} {\bibfield  {journal} {\bibinfo  {journal} {Science}\ }\textbf
  {\bibinfo {volume} {346}},\ \bibinfo {pages} {448} (\bibinfo {year}
  {2014})}\BibitemShut {NoStop}%
\bibitem [{\citenamefont {Wang}\ \emph
  {et~al.}(2015{\natexlab{b}})\citenamefont {Wang}, \citenamefont {Gao},
  \citenamefont {Wen}, \citenamefont {Han}, \citenamefont {Taniguchi},
  \citenamefont {Watanabe}, \citenamefont {Koshino}, \citenamefont {Hone},\
  and\ \citenamefont {Dean}}]{wang2015evidence}%
  \BibitemOpen
  \bibfield  {author} {\bibinfo {author} {\bibfnamefont {L.}~\bibnamefont
  {Wang}}, \bibinfo {author} {\bibfnamefont {Y.}~\bibnamefont {Gao}}, \bibinfo
  {author} {\bibfnamefont {B.}~\bibnamefont {Wen}}, \bibinfo {author}
  {\bibfnamefont {Z.}~\bibnamefont {Han}}, \bibinfo {author} {\bibfnamefont
  {T.}~\bibnamefont {Taniguchi}}, \bibinfo {author} {\bibfnamefont
  {K.}~\bibnamefont {Watanabe}}, \bibinfo {author} {\bibfnamefont
  {M.}~\bibnamefont {Koshino}}, \bibinfo {author} {\bibfnamefont
  {J.}~\bibnamefont {Hone}}, \ and\ \bibinfo {author} {\bibfnamefont {C.~R.}\
  \bibnamefont {Dean}},\ }\href@noop {} {\bibfield  {journal} {\bibinfo
  {journal} {Science}\ }\textbf {\bibinfo {volume} {350}},\ \bibinfo {pages}
  {1231} (\bibinfo {year} {2015}{\natexlab{b}})}\BibitemShut {NoStop}%
\bibitem [{\citenamefont {Kumar}\ \emph {et~al.}(2016)\citenamefont {Kumar},
  \citenamefont {Kuiri}, \citenamefont {Jung}, \citenamefont {Das},\ and\
  \citenamefont {Das}}]{kumar2016tunability}%
  \BibitemOpen
  \bibfield  {author} {\bibinfo {author} {\bibfnamefont {C.}~\bibnamefont
  {Kumar}}, \bibinfo {author} {\bibfnamefont {M.}~\bibnamefont {Kuiri}},
  \bibinfo {author} {\bibfnamefont {J.}~\bibnamefont {Jung}}, \bibinfo {author}
  {\bibfnamefont {T.}~\bibnamefont {Das}}, \ and\ \bibinfo {author}
  {\bibfnamefont {A.}~\bibnamefont {Das}},\ }\href@noop {} {\bibfield
  {journal} {\bibinfo  {journal} {Nano letters}\ }\textbf {\bibinfo {volume}
  {16}},\ \bibinfo {pages} {1042} (\bibinfo {year} {2016})}\BibitemShut
  {NoStop}%
\bibitem [{\citenamefont {Song}\ \emph {et~al.}(2015)\citenamefont {Song},
  \citenamefont {Samutpraphoot},\ and\ \citenamefont
  {Levitov}}]{song2015topological}%
  \BibitemOpen
  \bibfield  {author} {\bibinfo {author} {\bibfnamefont {J.~C.}\ \bibnamefont
  {Song}}, \bibinfo {author} {\bibfnamefont {P.}~\bibnamefont {Samutpraphoot}},
  \ and\ \bibinfo {author} {\bibfnamefont {L.~S.}\ \bibnamefont {Levitov}},\
  }\href@noop {} {\bibfield  {journal} {\bibinfo  {journal} {Proceedings of the
  National Academy of Sciences}\ }\textbf {\bibinfo {volume} {112}},\ \bibinfo
  {pages} {10879} (\bibinfo {year} {2015})}\BibitemShut {NoStop}%
\bibitem [{\citenamefont {Zhang}\ \emph {et~al.}(2005)\citenamefont {Zhang},
  \citenamefont {Tan}, \citenamefont {Stormer},\ and\ \citenamefont
  {Kim}}]{zhang2005experimental}%
  \BibitemOpen
  \bibfield  {author} {\bibinfo {author} {\bibfnamefont {Y.}~\bibnamefont
  {Zhang}}, \bibinfo {author} {\bibfnamefont {Y.-W.}\ \bibnamefont {Tan}},
  \bibinfo {author} {\bibfnamefont {H.~L.}\ \bibnamefont {Stormer}}, \ and\
  \bibinfo {author} {\bibfnamefont {P.}~\bibnamefont {Kim}},\ }\href@noop {}
  {\bibfield  {journal} {\bibinfo  {journal} {Nature}\ }\textbf {\bibinfo
  {volume} {438}},\ \bibinfo {pages} {201} (\bibinfo {year}
  {2005})}\BibitemShut {NoStop}%
\bibitem [{\citenamefont {Novoselov}\ \emph {et~al.}(2005)\citenamefont
  {Novoselov}, \citenamefont {Geim}, \citenamefont {Morozov}, \citenamefont
  {Jiang}, \citenamefont {Katsnelson}, \citenamefont {Grigorieva},
  \citenamefont {Dubonos},\ and\ \citenamefont {Firsov}}]{novoselov2005two}%
  \BibitemOpen
  \bibfield  {author} {\bibinfo {author} {\bibfnamefont {K.~S.}\ \bibnamefont
  {Novoselov}}, \bibinfo {author} {\bibfnamefont {A.~K.}\ \bibnamefont {Geim}},
  \bibinfo {author} {\bibfnamefont {S.}~\bibnamefont {Morozov}}, \bibinfo
  {author} {\bibfnamefont {D.}~\bibnamefont {Jiang}}, \bibinfo {author}
  {\bibfnamefont {M.}~\bibnamefont {Katsnelson}}, \bibinfo {author}
  {\bibfnamefont {I.}~\bibnamefont {Grigorieva}}, \bibinfo {author}
  {\bibfnamefont {S.}~\bibnamefont {Dubonos}}, \ and\ \bibinfo {author}
  {\bibfnamefont {A.}~\bibnamefont {Firsov}},\ }\href@noop {} {\bibfield
  {journal} {\bibinfo  {journal} {nature}\ }\textbf {\bibinfo {volume} {438}},\
  \bibinfo {pages} {197} (\bibinfo {year} {2005})}\BibitemShut {NoStop}%
\bibitem [{\citenamefont {Jung}\ \emph {et~al.}(2014)\citenamefont {Jung},
  \citenamefont {Raoux}, \citenamefont {Qiao},\ and\ \citenamefont
  {MacDonald}}]{jung2014ab}%
  \BibitemOpen
  \bibfield  {author} {\bibinfo {author} {\bibfnamefont {J.}~\bibnamefont
  {Jung}}, \bibinfo {author} {\bibfnamefont {A.}~\bibnamefont {Raoux}},
  \bibinfo {author} {\bibfnamefont {Z.}~\bibnamefont {Qiao}}, \ and\ \bibinfo
  {author} {\bibfnamefont {A.~H.}\ \bibnamefont {MacDonald}},\ }\href@noop {}
  {\bibfield  {journal} {\bibinfo  {journal} {Physical Review B}\ }\textbf
  {\bibinfo {volume} {89}},\ \bibinfo {pages} {205414} (\bibinfo {year}
  {2014})}\BibitemShut {NoStop}%
\bibitem [{\citenamefont {Altshuler}\ \emph {et~al.}(1982)\citenamefont
  {Altshuler}, \citenamefont {Aronov},\ and\ \citenamefont
  {Khmelnitsky}}]{altshuler1982effects}%
  \BibitemOpen
  \bibfield  {author} {\bibinfo {author} {\bibfnamefont {B.~L.}\ \bibnamefont
  {Altshuler}}, \bibinfo {author} {\bibfnamefont {A.}~\bibnamefont {Aronov}}, \
  and\ \bibinfo {author} {\bibfnamefont {D.}~\bibnamefont {Khmelnitsky}},\
  }\href@noop {} {\bibfield  {journal} {\bibinfo  {journal} {Journal of Physics
  C: Solid State Physics}\ }\textbf {\bibinfo {volume} {15}},\ \bibinfo {pages}
  {7367} (\bibinfo {year} {1982})}\BibitemShut {NoStop}%
\bibitem [{\citenamefont {Stauber}\ \emph {et~al.}(2007)\citenamefont
  {Stauber}, \citenamefont {Peres},\ and\ \citenamefont
  {Guinea}}]{stauber2007electronic}%
  \BibitemOpen
  \bibfield  {author} {\bibinfo {author} {\bibfnamefont {T.}~\bibnamefont
  {Stauber}}, \bibinfo {author} {\bibfnamefont {N.}~\bibnamefont {Peres}}, \
  and\ \bibinfo {author} {\bibfnamefont {F.}~\bibnamefont {Guinea}},\
  }\href@noop {} {\bibfield  {journal} {\bibinfo  {journal} {Physical Review
  B}\ }\textbf {\bibinfo {volume} {76}},\ \bibinfo {pages} {205423} (\bibinfo
  {year} {2007})}\BibitemShut {NoStop}%
\bibitem [{\citenamefont {Hwang}\ and\ \citenamefont
  {Sarma}(2008)}]{hwang2008acoustic}%
  \BibitemOpen
  \bibfield  {author} {\bibinfo {author} {\bibfnamefont {E.}~\bibnamefont
  {Hwang}}\ and\ \bibinfo {author} {\bibfnamefont {S.~D.}\ \bibnamefont
  {Sarma}},\ }\href@noop {} {\bibfield  {journal} {\bibinfo  {journal}
  {Physical Review B}\ }\textbf {\bibinfo {volume} {77}},\ \bibinfo {pages}
  {115449} (\bibinfo {year} {2008})}\BibitemShut {NoStop}%
\bibitem [{\citenamefont {Venugopal}\ \emph {et~al.}(2011)\citenamefont
  {Venugopal}, \citenamefont {Chan}, \citenamefont {Li}, \citenamefont
  {Magnuson}, \citenamefont {Kirk}, \citenamefont {Colombo}, \citenamefont
  {Ruoff},\ and\ \citenamefont {Vogel}}]{venugopal2011effective}%
  \BibitemOpen
  \bibfield  {author} {\bibinfo {author} {\bibfnamefont {A.}~\bibnamefont
  {Venugopal}}, \bibinfo {author} {\bibfnamefont {J.}~\bibnamefont {Chan}},
  \bibinfo {author} {\bibfnamefont {X.}~\bibnamefont {Li}}, \bibinfo {author}
  {\bibfnamefont {C.~W.}\ \bibnamefont {Magnuson}}, \bibinfo {author}
  {\bibfnamefont {W.~P.}\ \bibnamefont {Kirk}}, \bibinfo {author}
  {\bibfnamefont {L.}~\bibnamefont {Colombo}}, \bibinfo {author} {\bibfnamefont
  {R.~S.}\ \bibnamefont {Ruoff}}, \ and\ \bibinfo {author} {\bibfnamefont
  {E.~M.}\ \bibnamefont {Vogel}},\ }\href@noop {} {\bibfield  {journal}
  {\bibinfo  {journal} {Journal of Applied Physics}\ }\textbf {\bibinfo
  {volume} {109}},\ \bibinfo {pages} {104511} (\bibinfo {year}
  {2011})}\BibitemShut {NoStop}%
\end{thebibliography}%
\end{document}